\documentclass[11pt]{article}
\pdfoutput=1
\usepackage{jheppub}
\usepackage{amsmath,amssymb,amsfonts,graphicx,slashed,color,amsthm,mathtools,upgreek, enumerate, tensor, subfig, bbold}
\usepackage{arydshln}
\allowdisplaybreaks

\allowdisplaybreaks

\newcommand{\bra}[1]{\langle #1 |}
\newcommand{\ket}[1]{| #1 \rangle}

\DeclareMathOperator{\tr}{tr}

\usepackage{hyperref}
\usepackage[utf8]{inputenc}
\usepackage[titletoc]{appendix}
\numberwithin{equation}{section} 

\title{Quantum Complexity of Time Evolution with Chaotic Hamiltonians}

\author[a, b]{Vijay Balasubramanian}
\author[a]{\!, Matthew DeCross}
\author[a]{\!, Arjun Kar}
\author[a]{\!, Onkar Parrikar}

\affiliation[\,a]{David Rittenhouse Laboratory, University of Pennsylvania,\\
209 S.33rd Street, Philadelphia PA, 19104, U.S.A.}
\affiliation[\,b]{Theoretische Natuurkunde, Vrije Universiteit Brussel (VUB), and \\ International Solvay Institutes, Pleinlaan 2, B-1050 Brussels, Belgium.}

\abstract{
We study the quantum complexity of time evolution in large-$N$ chaotic systems, with the SYK model as our main example. This complexity is expected to increase linearly for exponential time prior to saturating at its maximum value, and is related to the length of minimal geodesics on the manifold of unitary operators that act on Hilbert space.  Using the Euler-Arnold formalism, we demonstrate that there is always a geodesic between the identity and the time evolution operator $e^{-iHt}$ whose length grows linearly with time.  This geodesic is minimal until there is an obstruction to its minimality, after which it can fail to be a minimum either locally or globally.  We identify a criterion -- the Eigenstate Complexity Hypothesis (ECH) -- which bounds the overlap between off-diagonal energy eigenstate projectors and the $k$-local operators of the theory, and use it to argue that the linear geodesic will at least be a local minimum for exponential time. We show numerically that the large-$N$ SYK model (which is chaotic) satisfies ECH and thus has no local obstructions to linear growth of complexity for exponential time, as expected from holographic duality. In contrast, we also study the case with $N=2$ fermions (which is integrable) and find short-time linear complexity growth followed by oscillations. Our analysis relates complexity to familiar properties of physical theories like their spectra and the structure of energy eigenstates and has implications for the hypothesized computational complexity class separations PSPACE $\nsubseteq$ BQP/poly and PSPACE $\nsubseteq$ BQSUBEXP/subexp, and the ``fast-forwarding'' of quantum Hamiltonians.

}

\keywords{}

\begin{document}

\maketitle

\parskip=10pt

\begin{section}{Introduction}
In recent years the late time dynamics of general relativity have been examined through various lenses.
Two of the most prominent directions in this subject deal with quantities whose classical behavior cannot possibly continue to hold into the asymptotic future due to fundamental quantum-mechanical obstructions. The first is the exponential decay of a CFT two-point function computed using classical gravity in an AdS black hole, which could break down at a time as early as $t\sim S$ (where $S$ is the entropy of the black hole) as a consequence of the unitarity of quantum mechanics \cite{Maldacena2003}.
The second is the linear growth of the Einstein-Rosen bridge in the two-sided eternal AdS wormhole geometry, which led to a conjecture relating bulk volume/action and boundary quantum circuit complexity \cite{Stanford:2014jda, Brown2016}.\footnote{Here ``circuit complexity" measures the minimum number of simple, perhaps locally acting, gates necessary to construct a desired state or operator from a fixed reference.} If this conjecture is correct, then the extrapolation of the gravity result to times beyond $t\sim e^S$ is expected to break down due to quantum effects in a finite-dimensional quantum gravity Hilbert subspace. Various studies of both quantum circuit complexity and correlation function behavior have explored these observations \cite{Brown2017, Cotler2017}. However, in the case of circuit complexity, despite the plethora of analytic results from gravity calculations (see \cite{Carmi:2017jqz, Fu:2018kcp, Agon:2018zso, Swingle2018a, Chapman2018a, Chapman2018b, Brown:2018bms, Caceres2019, Caceres:2018blh, Goto2019} and references therein) assuming the volume/action conjecture, there has been little non-perturbative progress towards a first principles calculation of circuit complexity in CFT. In this paper, we seek to remedy this situation by studying the complexity of time evolution with chaotic Hamiltonians (which are expected to have gravity duals), especially with an eye towards the late-time behavior. 

At present, the most accessible method to compute complexity in continuum quantum systems is Nielsen's geometric formulation \cite{Nielsen2003,Nielsen2005,Nielsen2006,Nielsen2007}.\footnote{But see \cite{Czech:2017ryf,Caputa:2017yrh} for proposed path integral approaches, which have not been shown to be polynomially equivalent to quantum circuit complexity.}
In this approach, the circuit complexity of a unitary operator $U$ is the length of the minimal geodesic on the unitary group joining the identity to $U$. One begins by classifying the Lie algebra of the unitary group into ``local'' or ``easy" directions, represented by operators $T_{\alpha}$, and ``non-local'' or ``hard" directions $T_{\dot{\alpha}}$. Typically, the local directions will consist of operators with less than $k$-body interactions, for some $k$. One then picks a right-invariant metric on the group $U$ with the appropriate cost factors built in, such that motion along hard directions is disincentivized. The geodesic length with such a metric was shown to be polynomially equivalent to the usual notion of circuit complexity, which involves counting elementary unitary gates, provided the cost factors are chosen to scale exponentially with the Hilbert space dimension \cite{Nielsen2006}.
Heuristically, one can think of the circuit as a sequence of gates which corresponds to a sequence of geodesic segments on the unitary manifold; the geodesic in the geometric framework is then an everywhere-smooth approximation to this piecewise-smooth curve (Fig.~\ref{fig:nielsen_geodesic}). In this work, we will be interested in this geodesic notion of complexity.

\begin{figure}
    \centering
    \includegraphics[scale=.75]{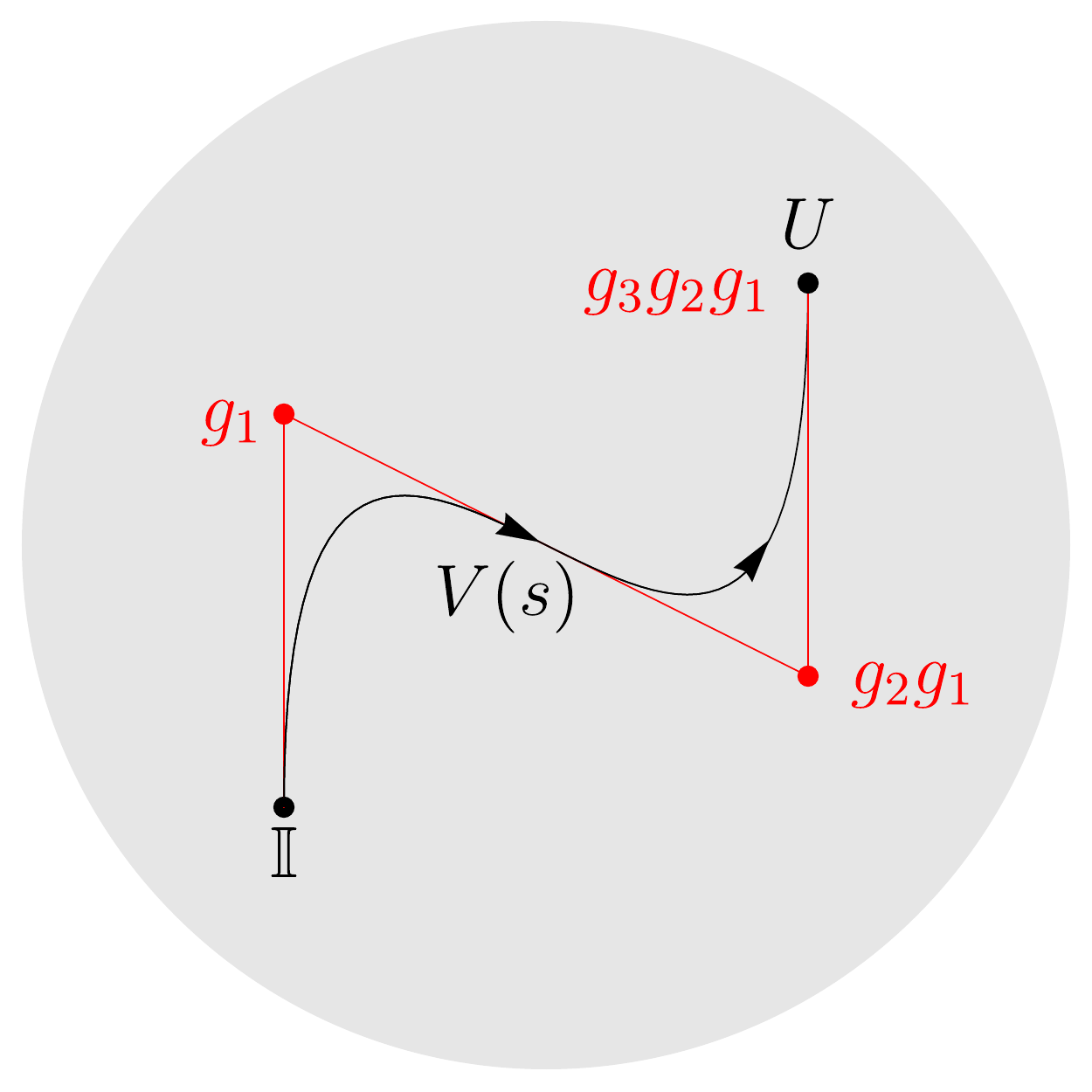}
    \caption{Schematic of the unitary manifold (gray disk).  A geodesic path (black) is depicted from the identity to some target unitary $U$.  The red straight lines represent construction of a circuit using some elementary gates $g_i$, and the final unitary is $U = g_3g_2g_1$.  The geodesic approximates the circuit smoothly by varying a control velocity $V(s)$, analogous to an infinitesimal elementary gate, where $s$ parametrizes the curve.}
    \label{fig:nielsen_geodesic}
\end{figure}

This technique has been applied by various authors to compute complexity in several physical systems \cite{Myers2017, Chapman:2017rqy, Khan:2018rzm, Myers2018} (see also \cite{Caputa:2017yrh, Hashimoto:2017fga, Kim:2017qrq, Moosa:2017yvt, Molina-Vilaplana:2018sfn, Alves:2018qfv, Magan:2018nmu, Caputa:2018kdj, Camargo:2018eof, Ali:2018fcz, Belin:2018bpg, Jiang:2018nzg} for related work, particularly on time evolution of complexity).
However, most applications so far have computed geodesics within a subspace of states or circuits, instead of dealing with the entire unitary group manifold. For instance, much recent work has focused on the subspace of Gaussian states, which are relevant in the context of free quantum field theories. 
This is because in continuum quantum-mechanical systems, the Hilbert space is often infinite-dimensional and it is difficult to define a tractable algebra of operators which generate the entire unitary group on the Hilbert space. Prior work which attempted to deal with the global structure of the unitary group relied on toy models \cite{Brown2017,Lin2018} of Lie group geometry. These models were constructed using metrics of strictly negative sectional curvature (or a discretization thereof, in the case of \cite{Lin2018}) in order to ensure chaotic behavior of geodesics on the unitary manifold \cite{Brown2017}. Here, we approach the problem of circuit complexity by  studying aspects of geodesics on the complete group manifold $SU(2^{N/2})$, which is the unitary group acting on the Hilbert space of $N/2$ qubits. Our primary motivation is to study complexity growth in chaotic quantum systems as opposed to free field theories. To this end, we will use the (generalized) Sachdev-Ye-Kitaev (SYK) model as a specific example of a chaotic Hamiltonian, although most of our arguments are general and should apply to any chaotic system.

Recall that the SYK model is a quantum-mechanical system comprising $N$ Majorana fermions $\psi_i$ with the Hamiltonian
\begin{equation}
H = \sum_{i_1 < \dots < i_q} J_{i_1 \dots i_q} \psi_{i_1} \dots \psi_{i_q} ,
\end{equation}
where the couplings $J_{i_1 \dots i_q}$ are drawn at random from a Gaussian distribution with mean zero and variance $\sigma^2$
\begin{equation}
\sigma^2 = \frac{(q-1)! \mathcal{J}^2}{N^{q-1}} ,
\end{equation}
where $\mathcal{J}$ is a parameter setting the variance \cite{Kitaev2015}.
This model is expected to be chaotic and holographically dual to 2D quantum gravity \cite{Maldacena2016, Polchinski:2016xgd, Kitaev:2017awl, Stanford:2017thb} (see also \cite{Sarosi:2017ykf} for a review and additional references). From the SYK perspective, the group $SU(2^{N/2})$ is the group of unitary operations (modulo an overall phase) acting on the Hilbert space of the $N$ Majorana fermions (with $N$ even) $\psi_i$. Our main tool in studying the complexity in this model will be the Euler-Arnold equation \cite{Arnold1966,Tao2010,Nielsen2007}, which was also used in a simpler setting in \cite{Balasubramanian2018}.

\begin{figure}[h!]
\begin{center}
\includegraphics[height=5.5cm]{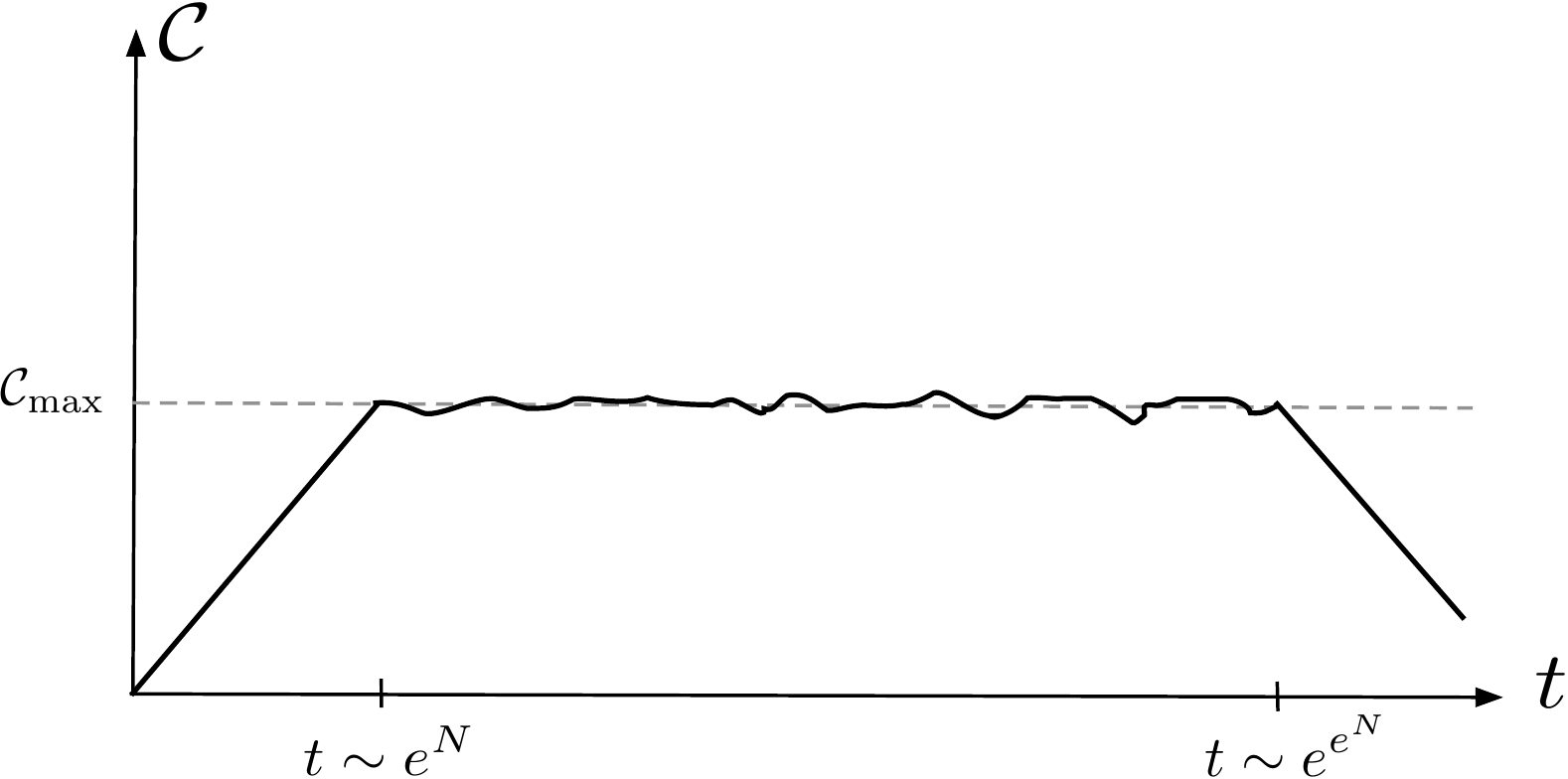} \caption{\small{The complexity in chaotic systems is conjectured \cite{Brown2017} to grow linearly in time until a time of order $e^N$, after which it saturates to (and fluctuates around) its maximum value of $\mathcal{C}_{\text{max}}$. At doubly exponential time, the complexity is expected to exhibit recurrences. }. \label{fig:CExp}}
\end{center}
\end{figure}

From physical considerations and holographic as well as complexity-theoretic arguments, the complexity in chaotic systems has been conjectured \cite{Brown2017} to grow linearly in time until a time of order $e^N$, after which it is expected to saturate to (and fluctuate around) its maximum value of $\mathcal{C}_{\text{max}} \sim \text{poly}(N) e^N$ (see Fig.~\ref{fig:CExp}), where by poly$(N)$ we mean $N^\alpha$ for some $\alpha \geq 0$. Here $N$ is the number of fermions in the SYK model, but more generally it should be taken to be log of the dimension of the Hilbert space. The motivation of the present work is to better understand the origin of this behavior and the various time scales involved from a field theory perspective, within the geodesic complexity framework. One of our main results will be to establish the existence and \emph{local minimality} of a geodesic between the identity and $e^{-iHt}$ whose length grows linearly with time $t$. The existence of such a geodesic only relies on general features such as the Hamiltonian being local (i.e., it should be built from easy generators), and uniformity of the cost factor in the easy directions. However, this is not the whole story -- the linear geodesic is not guaranteed to be a local minimum of the distance function (i.e., it could be a saddle point), much less a global minimum. As such, it may not be the relevant geodesic for complexity. In this paper, we will investigate in depth the question of local minimality of the linear geodesic by studying \emph{conjugate points} along it. Roughly, we say that we have a conjugate point at time $t$ if we can deviate infinitesimally from the linear geodesic at time $t=0$ (i.e., deform the initial velocity infinitesimally) and return to it at time $t$ along an infinitesimally nearby curve which satisfies the geodesic equation linearized to first order. The original geodesic stops being minimizing past the first conjugate point (i.e., it is a saddle point thereafter), and so for the physical considerations explained in Fig.~\ref{fig:CExp} to be correct, it is necessary (but not sufficient) that the no conjugate points appear along the linear geodesic at times sub-exponential in $N$. We will give an argument that this is indeed the case for ``sufficiently chaotic" Hamiltonians (such as the SYK model) and for an appropriate choice of the cost factors. Therefore, the linear geodesic is at least locally minimizing for times exponential in $N$, consistent with the expectations in Fig.~\ref{fig:CExp}. Our proof will involve a new criterion on the Hamiltonian from the vantage point of circuit complexity which we will call the \emph{eigenstate complexity hypothesis} (ECH):

\noindent\textbf{Eigenstate Complexity Hypothesis (ECH)}: Let $H$ be the Hamiltonian with energy eigenstates $|m\rangle, |n\rangle$ etc., $T_{\alpha}$ be the local generators in the Lie algebra, and $T_{\dot\alpha}$ be the non-local generators. Define
\begin{equation}
    R_{mn} = \frac{\sum_{\alpha} |\langle m|T_{\alpha}|n\rangle |^2}{\sum_{\alpha} |\langle m|T_{\alpha}|n\rangle |^2+\sum_{\dot\alpha} |\langle m|T_{\dot\alpha}|n\rangle |^2}.
\end{equation}
We will say that the Hamiltonian and the gate set satisfy the eigenstate complexity hypothesis, if for $E_m\neq E_n$ in the large-$N$ limit,
\begin{equation}
    R_{mn} = e^{-2S} \text{poly}(S)\, r_{mn},\label{eqECH}
\end{equation}
where $S$ is the log dimension of the Hilbert space (i.e., $\frac{N}{2}\ln\,2$ for the SYK model) and $r_{mn}$ are $O(1)$ numbers which do not scale with $S$.

In words, ECH is the condition that off-diagonal eigenstate projectors of the form $\ket{m}\bra{n}$ which map one energy eigenstate of the Hamiltonian to a different eigenstate should have $e^{-S}$ suppressed overlaps with the easy/local/simple directions in the gate set, or equivalently, such off-diagonal energy eigenstate projectors must necessarily be ``complex'' (i.e., complicated).\footnote{We discuss the relationship with the well-known Eigenstate Thermalization Hypothesis (ETH) \cite{PhysRevA.43.2046, 1994PhRvE..50..888S} in the main text.} For Hamiltonians which satisfy the ECH, the conjugate point analysis simplifies greatly, and the exponential bound on conjugate points can be analytically argued. We will provide numerical evidence to show that the SYK model indeed satisfies the ECH.

The rest of this paper is organized as follows:
in section \ref{sec:SU(N)}, we will begin by briefly reviewing the geodesic complexity framework and setting up the Euler-Arnold formalism for studying the complexity of local Hamiltonians in the Lie algebra $\mathfrak{su}(2^{N/2})$ for even $N$. In sections \ref{sec:N=2analytics} and \ref{sec:numericsN=2}, we study the simple case of $N=2$ where all the geodesics between identity and $e^{-iHt}$ can be worked out and the complexity calculated using analytic and numerical techniques. In section \ref{sec:generalNlin}, we will switch to general $N$ and show the existence of a geodesic whose length grows linearly with time. In section \ref{sec:conjugate}, we will explore the local minimality of the linear geodesic by studying conjugate points. We will end with some remarks on late-time saturation, complexity classes, and quantum chaos in the Discussion (section \ref{sec:discussion}).  
\end{section}

\begin{section}{Geometry of $SU(2^{N/2}$)}{\label{sec:SU(N)}}
The Hilbert space of $N/2$ qubits has a natural tensor factorization
\begin{equation}
\mathcal{H} = \underbrace{\mathbb{C}^2 \otimes \dots \otimes \mathbb{C}^2}_{N/2} .
\end{equation}
We wish to study the geometry of the set of (unit-determinant) unitary operators $\mathcal{U}(\mathcal{H})$ that acts on this Hilbert space.
In this case, this set is
\begin{equation}
\mathcal{U}(\mathcal{H}) = \mathcal{U}(\mathbb{C}^{2^{N/2}}) = SU(2^{N/2}) .
\label{eq:hilbertspace}
\end{equation}
In order to study the differential geometry of $SU(2^{N/2})$ from the quantum computation viewpoint, we must pick a basis for the Lie algebra with some notion of locality, i.e., we should be able to identify some generators in the Lie algebra as local or ``simple'', and the rest as ``complex''. In quantum computation, we usually choose some simple unitary operators as the elementary gates to be used in building circuits. On the other hand, in the geodesic framework, it is natural to choose a $k$-local subspace of the Lie algebra of the unitary group manifold to correspond to ``simple directions''. We may think of the elementary gates of the quantum computation viewpoint as being exponentials of these simple generators.
For general $\mathcal{H}$, there is no guarantee that we can choose a basis for the unitary Lie algebra which respects any sort of locality.
Luckily, for the qubit case $SU(2^{N/2})$, there are a couple of natural ways to proceed. We could pick the ``Pauli basis'', namely products of Pauli matrices acting on individual qubits, as our basis of generators. However, there is a second choice which is more natural from the point of view of the SYK model: consider the gamma matrices $\gamma_a$ with $a \in \{0, \dots , N-1\}$ which satisfy the Clifford algebra (with $\gamma_a^\dagger = \gamma_a$):
\begin{equation}
\{ \gamma_a , \gamma_b\} = 2\delta_{ab} .
\label{eq:cliffordalg}
\end{equation}
Now consider distinct ordered products $T_{a_1\cdots a_m}=\gamma_{a_1} \dots \gamma_{a_m}$ with $m \in \{1, \dots , N\}$ and $a_p < a_q$ for $p < q$. We will often denote these operators as simply $T_{i}$, where $i$ stands for the multi-index $a_1\cdots a_m$. The total number of such ordered products is $\sum_{m=1}^N {{N}\choose{m}} = 2^{N}-1$.
This is precisely the dimension of the Lie algebra $\mathfrak{su}(2^{N/2})$.
\end{section}
It is simple to make such ordered products of gamma matrices Hermitian by inserting appropriate factors of $i$. 
We claim this construction is a basis for $\mathfrak{su}(2^{N/2})$, and we leave the proof to appendix \ref{sec:basis}.
We can endow the gamma matrix basis with a natural notion of locality as follows: $k$-local generators of the Lie algebra are simply those involving $k$ or fewer gamma matrices.
This is precisely the natural notion of locality in the SYK model -- from this point of view, the gamma matrices above correspond to the Majorana fermion operators $\psi_a$ in the SYK model. 

The basic idea in the geodesic framework is to model circuit complexity \cite{Nielsen2007} in terms of a minimal-length geodesic on $SU(2^{N/2})$ with respect to a right-invariant metric chosen such that it disincentivizes motion in the directions of nonlocal unitary operators.
This corresponds to a choice of gate set in the quantum computation picture, where we allow up to $k$-local gates (i.e., exponentials of $k$-local generators in the Lie algebra) in our circuit but do not allow more nonlocal gates.
In our context, we want to disincentivize motion in directions which correspond to generators involving products of more than $k$ gamma matrices.
Let us begin by constructing such a right-invariant metric.
We can use the gamma matrix basis for $\mathfrak{su}(2^{N/2})$ to compute the structure constants ${f_{ij}}^\ell$ of the Lie algebra, defined as\footnote{When sums are not written explicitly, the Einstein summation convention is adopted.  We caution the reader that there will be expressions in which repeated indices appear three times, but this will not cause any ambiguities because the three matching indices will always be summed over together.}
\begin{equation}
[T_i , T_j ] = i {f_{ij}}^\ell T_\ell ,
\label{eq:liealgebra}
\end{equation}
where recall that the $T_i= \gamma_{a_1}\cdots\gamma_{a_m}$ are generators built from products of gamma matrices (or equivalently, products of the SYK fermion operators) labelled by the multi-index $i=(a_1\cdots a_m)$. Using these, we calculate the Cartan-Killing form
\begin{equation}
K_{ij} = -\frac{1}{h^\vee} {f_{im}}^\ell {f_{j\ell}}^m ,
\label{eq:cartankilling}
\end{equation}
(where $h^\vee$ is the dual Coxeter number) which is a positive-definite\footnote{Some definitions of the Cartan-Killing form instead yield a negative-definite form for compact Lie algebras.  We are only interested in this form up to overall sign and normalization since our only use for it is to define a right-invariant Riemannian metric on $SU(2^{N/2})$.} bilinear form. In order to build in the notion of simple and hard directions in the Lie algebra, we construct a new positive-definite bilinear form on $\mathfrak{su}(2^{N/2})$
\begin{equation}
G_{ij} = \frac{c_i+c_j}{2} K_{ij},
\label{eq:algmetric}
\end{equation}
where the numbers $c_i$ are ``cost factors".
Then a right-invariant metric $g$ can be defined at an arbitrary point $U$ on $SU(2^{N/2})$ by simply taking
\begin{equation}
g_U(X,Y) = G(XU^{-1},YU^{-1}) ,
\label{eq:groupmetric}
\end{equation}
where we have used the group structure to transport the tangent vectors $X$ and $Y$ from $U$ back to the identity and then applied \eqref{eq:algmetric}.
The cost factors $c_i$ encode the information about our choice of local and nonlocal directions, i.e. our notion of $k$-locality.
We will generally take $c_i = 1$ if the generator $T_i$ consists of $k$ or fewer gamma matrices, and $c_i= 1+\mu$ with $\mu\gg 1$ otherwise; we will specify how large $\mu$ has to be shortly. 
Note that if we chose cost factors $c_i = 1$ for all $i$, the metric \eqref{eq:groupmetric} would actually be {\it bi-invariant}. Here bi-invariant means that the metric is both left and right invariant. The restriction to right-invariance arises by choosing at least one cost factor to be $c_i \neq 1$ (or more generally by choosing a symmetric bilinear form for the metric which is not proportional to the identity).

Having chosen our cost factors, the geodesic equation on $SU(2^{N/2})$ with metric \eqref{eq:groupmetric} is given in terms of the Lie algebra metric \eqref{eq:algmetric} and structure constants \eqref{eq:liealgebra} by the Euler-Arnold equation \cite{Arnold1966}\footnote{The original article is in French, but an English summary can be found in \cite{Tao2010}.}
\begin{equation}
G_{ij} \frac{dV^j}{ds} = {f_{ij}}^p V^j G_{p\ell} V^\ell ,
\label{eq:eulerarnold}
\end{equation}
where the velocities $V^i(s)$ control the unitary path the geodesic follows via
\begin{equation}
U(s) = \mathcal{P} \exp \left( -i \int_0^s ds' V^i(s') T_i \right) ,
\label{eq:path-ordered-solution}
\end{equation}
and we have made use of the path-ordered exponential to solve the matrix equation for the unitary operator
\begin{equation}
\frac{dU}{ds} = -i V^i(s) T_i U(s) .
\label{eq:unitarydiffeq}
\end{equation}
Finally, we impose the boundary condition
\begin{equation}
U(1) = U_{\text{target}}
\end{equation}
for some target unitary whose circuit complexity we wish to study. This complexity is given by the geodesic length
\begin{equation}
\mathcal{C}(U_{\text{target}}) = \text{min}\,\int_0^1 ds \sqrt{G_{ij} V^i(s) V^j(s)} 
\end{equation}
where the minimization is over all geodesics from the identity to $U_{\text{target}}$. Throughout this paper, we will be interested in $U_{\text{target}} = e^{-iHt}$, where $H$ is a suitable $k$-local Hamiltonian.

Before moving on to calculations, we would like to specify how large the cost factor $\mu$ needs to be. The essential reason for choosing $\mu$ large is that it prevents the geodesic from wandering off in the hard, i.e., non-$k$-local directions, and we can then take such a geodesic to be a reasonable approximation to the true minimal circuit built only out of the allowed $k$-local gates (more precisely, gates of the form $g_{\alpha} = e^{i\epsilon T_{\alpha}}$ for $k$-local $T_{\alpha}$). Now imagine that we start with our minimal geodesic $U(s)$, and then define a new curve $\widetilde{U}(s)$ by simply projecting out from its velocity $V(s)$ all the hard directions. Having done so, this new curve would deviate from the target unitary by some amount, which we would like to be smaller than some fixed error tolerance. It was shown in \cite{Nielsen2006} that\footnote{In lemma 1 of \cite{Nielsen2006} a weaker inequality was proven with the coefficient of the right hand side being $e^S$ as opposed to $e^{S/2}$. However, one can do better by using the fact that the operator norm is upper bounded by the Frobenius norm in step 3 of the derivation found in the appendix of \cite{Nielsen2006}.}
\begin{equation}
|| U_{\text{target}} - \widetilde{U}(1) ||_{\text{op.}} \leq \frac{e^{S/2}}{\sqrt{1+\mu}}\mathcal{C}(U_{\text{target}}).
\end{equation}
If we pick $1+\mu = \frac{1}{\epsilon^2}e^{S}\mathcal{C}^2$, then the right hand side above can be made smaller than $\epsilon$. Therefore, for any target unitary with polynomial complexity\footnote{Our arguments should work more generally for any target unitary with sub-exponential complexity. For e.g., when $t$ scales sub-exponentially with $S$, then $e^{-itH}$ necessarily has at best sub-exponential complexity, as will become clear later.}, we need to take the cost factor to be $\mu \propto e^{S}$, where the proportionality factor may scale at most polynomially in $S$. 
\subsection{Analytics for $N=2$}
\label{sec:N=2analytics}
We will mainly be interested in studying the complexity for a large-$N$ chaotic Hamiltonian, with the SYK model as a specific example. However, as a warm up, we will begin with the case of an SYK-like model with $N=2$ fermions. The algebra for the $N=2$ case is simply the familiar $\mathfrak{su}(2)$ (see also \cite{Brown:2019whu}).
There are three generators, built from the Hermitian matrices $\gamma_a$.
\begin{equation}
\begin{split}
T_1 & = \gamma_1 , \\
T_2 & = \gamma_2 , \\
T_3 & \equiv T_{12} = i \gamma_1 \gamma_2 .
\end{split}
\end{equation}
We can compute the structure constants by using the algebra \eqref{eq:cliffordalg}.
\begin{align}
\begin{split}
[T_1,T_2] & = -2 i T_3 , \\
[T_2,T_3] & = -2 i T_1 ,\\
[T_3,T_1] & = -2 i T_2 .
\end{split}
\end{align}
We see that, even though we have chosen a slightly unusual basis for the algebra, the structure constants are still ${f_{ij}}^k = -2\epsilon_{ij\ell}\delta^{\ell k}$; this is essentially the usual angular momentum algebra up to a minus sign and a factor of 2.
This fact will allow us to solve the Euler-Arnold equation directly.
The Cartan-Killing form is given by\footnote{We will always normalize the Cartan-Killing form to $\delta_{ij}$, regardless of the coefficient obtained by using \eqref{eq:cartankilling}.}
\begin{equation}
K_{ij} = \delta_{ij}.
\end{equation}
Let us pick $c_1 = c_2 = 1$ and $c_3 = 1+\mu$, where $\mu$ is a large suppression factor to discourage motion in the $T_3$ direction.
This corresponds to enforcing $k=1$ locality.
The equations \eqref{eq:eulerarnold} then reduce to
\begin{equation}
\begin{split}
\frac{dV^1}{ds} & = -2 \mu V^2 V^3 ,\\
\frac{dV^2}{ds} & = 2 \mu V^3 V^1 ,\\
(1+\mu) \frac{dV^3}{ds} & = 0 ,
\end{split}
\end{equation}
and this system can be solved to find the unique solution with integration constants $v^i$.
\begin{equation}
\begin{split}
V^1(s) & = v^1 \cos \left( v^3 \mu s \right) - v^2 \sin \left( v^3 \mu s \right) , \\
V^2(s) & = v^2 \cos \left( v^3 \mu s \right) + v^1 \sin \left( v^3 \mu s \right) , \\
V^3(s) & = v^3 /2 .
\end{split}
\end{equation}
Thus far we have solved the geodesic equation at the level of the Lie algebra, which allows us to obtain the tangent vector at any point along the geodesic given an initial direction.
In fact, we can already compute the complexity of a path connecting $U(0) = 1$ to $U(1) = U_{\text{target}}$:
\begin{align}
\begin{split}
\mathcal{C} & = \int_0^1 ds \sqrt{(V^1)^2 + (V^2)^2 + (1+\mu) (V^3)^2} \\
& = \sqrt{(v^1)^2 + (v^2)^2 + \frac{1+\mu}{4} (v^3)^2} . \label{eq:complexity}
\end{split}
\end{align}
We see that the integrand is actually independent of $s$, leading to a simple result.
All the information about the path length is contained in the magnitude of the tangent vector at the identity.

We really would like to know the geodesic for fixed boundary conditions $U(0) = 1$ and $U(1) = U_{\text{target}}$ in order to fix the initial tangent vector $v^iT_i$.
The unitary $U(s)$ along the geodesic path from the identity with tangent vector $V^i(s) T_i$ is given by the path-ordered exponential \eqref{eq:path-ordered-solution}.
Now, we want to explicitly evaluate what the final unitary $U(1)$ looks like as a function of the initial velocity $v^i$, and then implement the boundary condition $U(1) = e^{-iHt}$ for some local, Hermitian Hamiltonian, in order to solve for $v^i$.
However, solving this would require us to brute-force deal with the path-ordering in \eqref{eq:path-ordered-solution}, which is a famously difficult problem and is solved in quantum mechanics (where there is a time-ordering rather than a path-ordering) using perturbation theory.
We would like a more nonperturbative approach, and we might hope that one exists since we are only dealing with finite-dimensional matrices rather than the infinite-dimensional Hilbert spaces familiar from other quantum systems like the harmonic oscillator.
Indeed, such a nonperturbative method for finite-dimensional matrix equations was found in \cite{Giscard2015}.
We employ their construction here.
Given the velocity along the geodesic
\begin{equation}
V(s) = V^i(s) T_i =  \left(
\begin{array}{cc}
 -v^3/2 & (v^1- i v^2) e^{-i v^3 \mu s}  \\
 (v^1+ i v^2) e^{i v^3 \mu s} & v^3/2 \\
\end{array}
\right) ,
\end{equation}
we wish to solve \eqref{eq:unitarydiffeq} subject to $U(0) = 1$ {\it without} the use of the path-ordering $\mathcal{P}$.
Let us define the frequency
\begin{equation}
\omega^2 = (v^1)^2 + (v^2)^2 + \frac{1}{4} (1+\mu)^2(v^3)^2 ,
\end{equation}
and the function
\begin{equation}
\varphi(s) = e^{-\frac{i}{2} v^3 \mu s} ,
\end{equation}
then the solution is
\begin{equation}
\label{eq:su2geodesic}
U(s) = \left(
\begin{array}{cc}
 \varphi(s) \left( \cos \omega s + i \frac{v^3(1+\mu)}{2\omega} \sin \omega s \right) &  -i \varphi(s) \frac{ (v^1 - i v^2)}{\omega} \sin \omega s \\
 -i \varphi(-s) \frac{ (v^1 + i v^2)}{\omega} \sin \omega s & \varphi(-s) \left(\cos \omega s - i \frac{v^3(1+\mu)}{2\omega} \sin\omega s \right)  \\
\end{array}
\right) .
\end{equation}
Note that this is a completely coordinate-free description of a path on the unitary manifold $SU(2)$; although $SU(2)$ happens to have a convenient interpretation as $S^3$, the higher groups $SU(2^{N/2})$ are nontrivial fiber bundles, so a coordinate patch-based method is likely difficult to implement.

We can now solve for $v^i$ (and hence compute the complexity of time evolution) by implementing the boundary condition $U(1) = e^{-i H t}$ for some Hamiltonian $H$, which we decompose as $H = \sum_i J_i T_i$.
The time evolution operator can be exactly computed with a simple matrix exponential, yielding (letting $J = \sqrt{J_1^2 + J_2^3 + J_3^2}$)
\begin{equation}
e^{-iHt} = \left(
\begin{array}{cc}
 \cos J t + i \frac{J_3}{J} \sin J t & -i\frac{J_1 - i J_2}{J} \sin J t \\
-i \frac{J_1 + i J_2}{J} \sin J t & \cos J t - i \frac{J_3}{J} \sin J t
\end{array}
\right).
\label{eq:timeevol}
\end{equation}
We can easily see that, if we had chosen all the metric cost factors to be $c_i = 1$ (i.e. taken $\mu = 0$ in \eqref{eq:su2geodesic}), the time evolution operator would itself define a geodesic curve.
This is because, for bi-invariant metrics, the matrix exponential coincides with the (Riemannian) exponential map.
In the next section, we will solve the boundary condition
\begin{equation}
\label{eq:boundarycondition}
U(1) = e^{-iHt} 
\end{equation}
for the velocities $v^i$ in terms of the Hamiltonian couplings $J_i$, for each value of $t$, using numerical techniques.
There will be, in general, multiple solutions to any such equation which correspond to different geodesics in $SU(2)$ which begin at the identity and end at $e^{-iHt}$.
We must of course find the one with minimal complexity.

\subsection{Numerics for $N=2$}
\label{sec:numericsN=2}
\begin{figure}[hbtp!]
\begin{center}
\includegraphics[width=.8\textwidth]{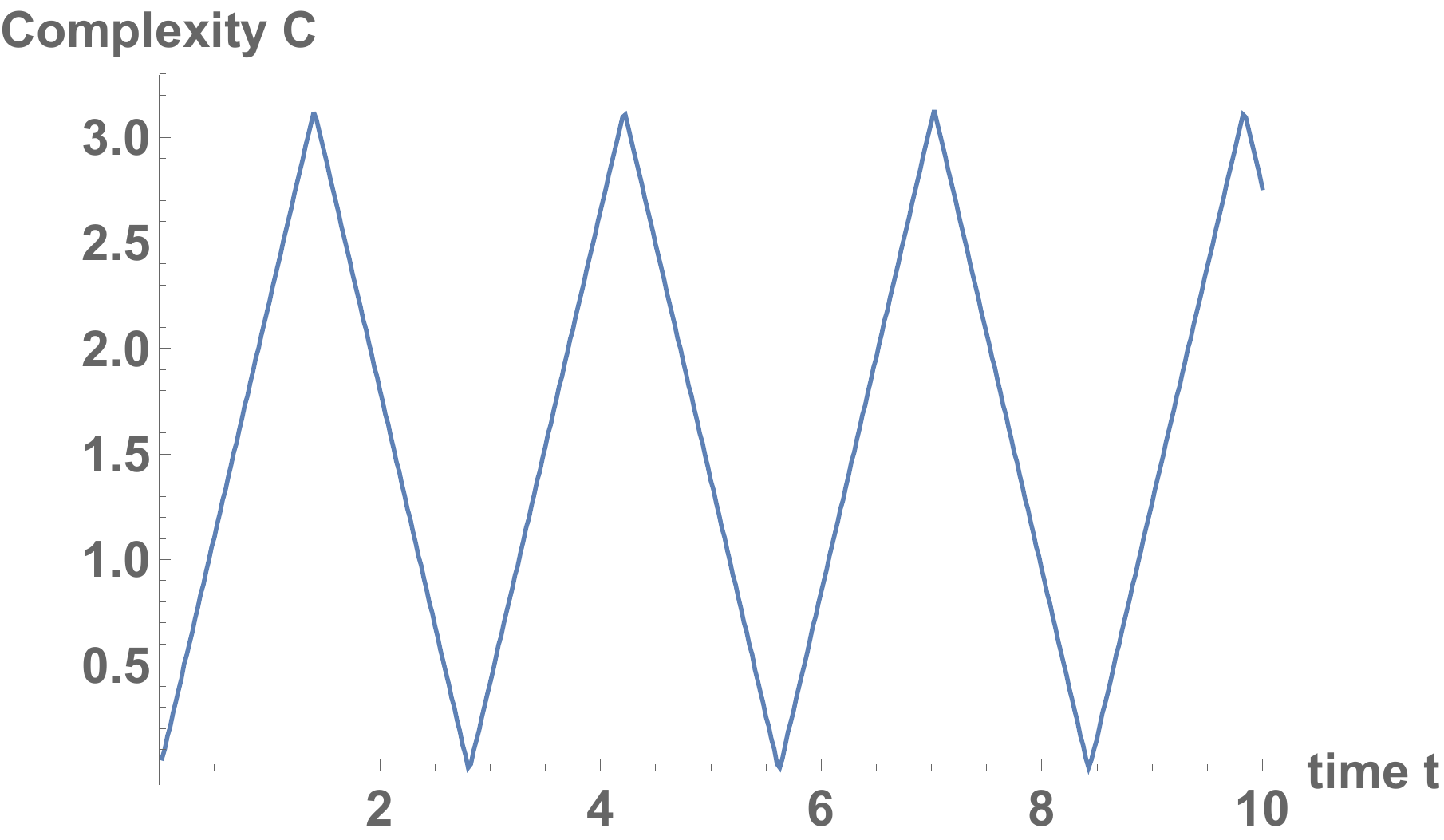}\caption{Complexity over time in appropriate dimensionless units with sample parameters $J_1 = 1$, $J_2 = 2$, $J_3 = 0$, $\frac{1}{1+\mu} = 0.09$. The complexity demonstrates an initial linear growth with the slope $J =  \sqrt{J_1^2 + J_2^2+J_3^2}$, attaining a maximum value of $\pi$, followed by linear oscillations (with slopes $\pm J$). \label{fig:numcomplex}}
\end{center}
\end{figure}

The matrix equation \eqref{eq:boundarycondition} reduces to a system of three independent transcendental equations in the velocities $v^1, v^2, v^3$. These can be solved numerically by brute force. Choosing the Hamiltonian to consist only of easy ($1$-local) operators, we work in the special case $J_3 = 0$. The numerical solution for the complexity \eqref{eq:complexity} as a function of time $t$ is presented in Fig.~\ref{fig:numcomplex}.

\begin{figure}[hbtp!]
\begin{center}
\includegraphics[width=\textwidth]{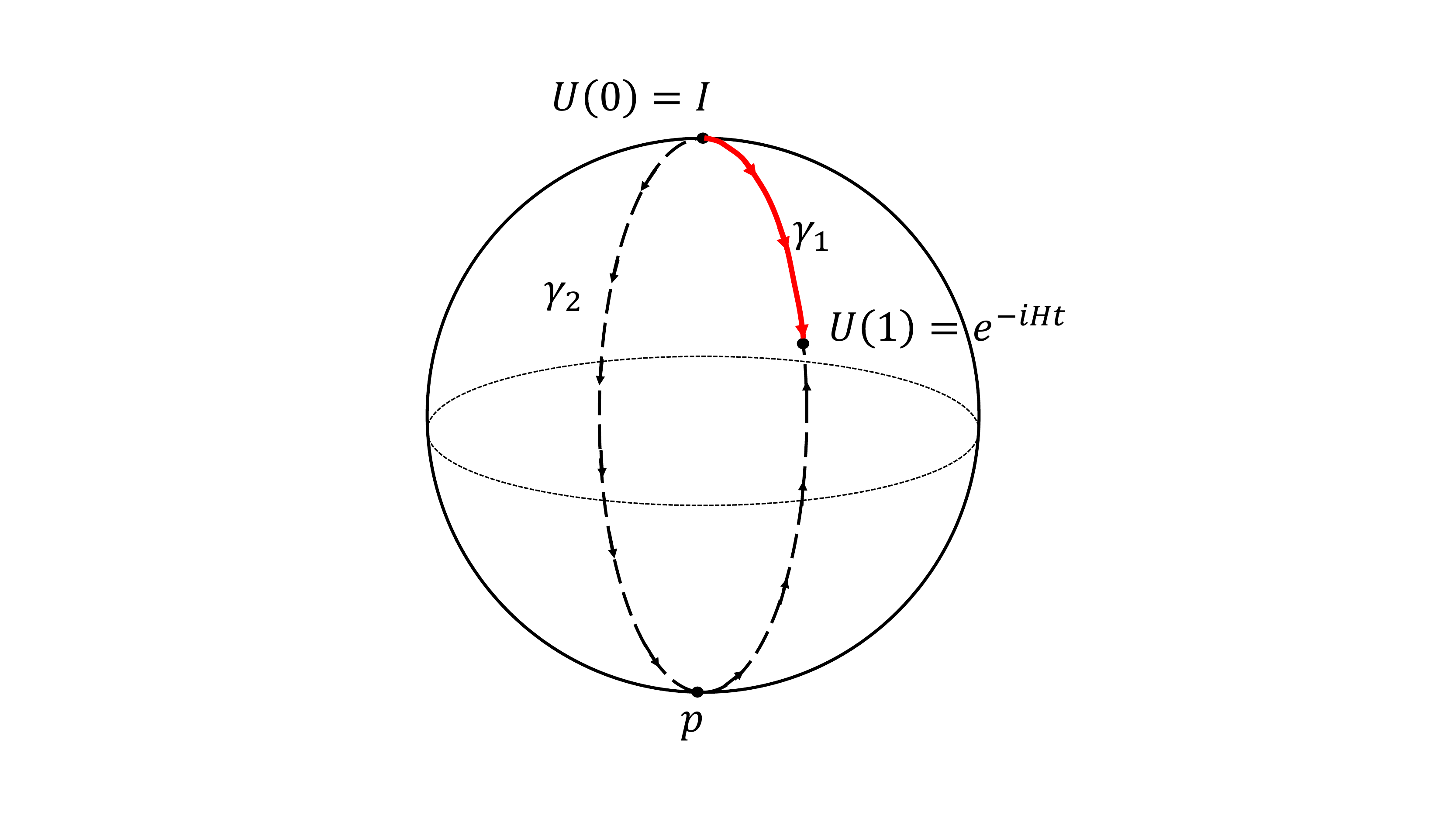} \caption{The geodesic $\gamma_1$ (red) lies on a great circle of $S^2$, connecting $U(0)$ and $U(1)$. At the antipodal point $p$, the geodesic $\gamma_2$ oriented oppositely along the same great circle exchanges dominance with $\gamma_1$. This effect leads to the linear decrease in complexity in $S^3$, i.e. in $SU(2)$. \label{fig:twosphere}}
\end{center}
\end{figure}

At early times, we find the expected linear growth of complexity, with slope $J   =  \sqrt{J_1^2 + J_2^2+J_3^2}$. At later times, however, one does not see the plateau predicted from holographic considerations; rather, there is an immediate linear decay of the complexity that may be considered to be a Poincar\'{e} recurrence. This behavior may be attributed to the simplicity of the group manifold $SU(2) \simeq S^3$ and may be visualized most easily on the two-sphere (Fig.~\ref{fig:twosphere}). 
After a certain maximum distance from the identity, the minimal length path on the sphere switches direction, i.e. the velocities $v_1, v_2$ change in sign. This maximum distance is equal to $\pi$, the maximum of the complexity in Fig.~\ref{fig:numcomplex}, as follows from results in \cite{su2dist}.

In the higher groups $SU(2^{N/2})$, we expect a quite nontrivial topology that results in a plateau in the complexity as many different geodesics exchange dominance at late enough times. In the simple case of $N=2$ fermions, we obviously do not see this effect, because the geodesic does not have sufficient space on the unitary group to wander around, away from its starting point. However, if we disorder average over the couplings $J_{1,2}$ in the Hamiltonian, then this is equivalent to considering an ensemble of systems where after the initial linear growth, we would expect to find ``cancellations'' between the various oscillating geodesic distances, thus leading to a plateau. In the SYK model, the couplings $J$ are drawn from a Gaussian distribution of zero mean with a variance chosen to simplify the large-$N$ limit, and the disorder average is performed during evaluation of correlation functions \cite{Maldacena2016}. Here we will simply disorder average the complexity directly, taking the couplings in the Hamiltonian $J_1, J_2$ to be drawn from Gaussian distributions of mean zero and some variance $\sigma^2$.

\begin{figure}[hbtp!]
\begin{center}
\includegraphics[width=.8\textwidth]{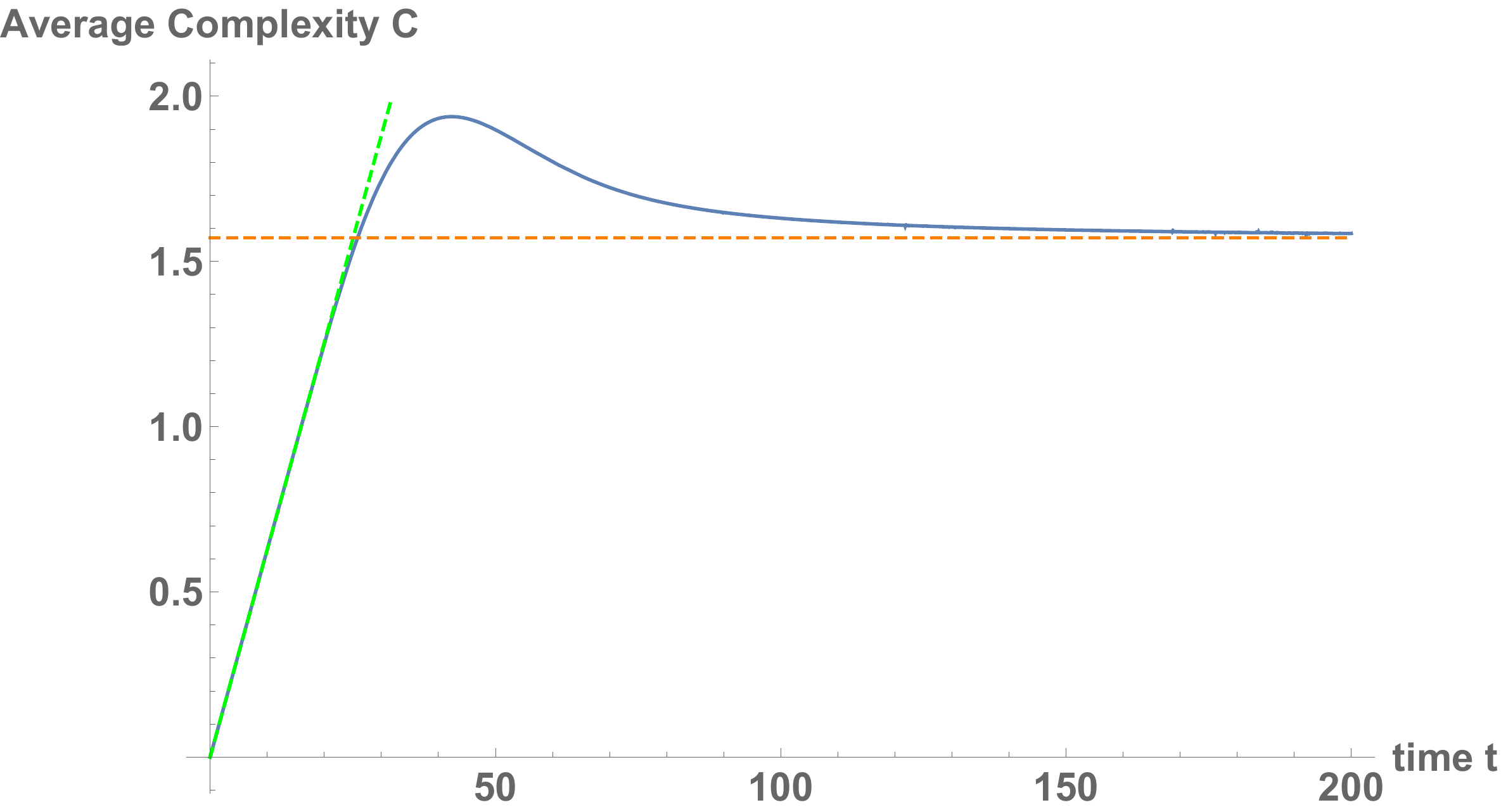} \caption{Disorder-averaged complexity (blue) as a function of time. At early times the complexity grows linearly with slope $\sqrt{\frac{\pi}{2}} \sigma$ (green), while at late times it approaches the asymptotic value $\pi/2$ (orange). \label{fig:complexaverage}}
\end{center}
\end{figure}

The result of this averaging procedure is displayed in Fig.~\ref{fig:complexaverage}. Specifically, we are plotting the disorder-averaged complexity:
\begin{align}
    \bar{C}(t) = \int_0^{\infty} dJ \: f_{\triangle} (t) \frac{J}{\sigma^2} e^{-\frac{J^2}{2\sigma^2}} \label{eq:averagedcomplex}
\end{align}
where $f_{\triangle} (t)$ refers to the triangle wave plotted in Fig.~\ref{fig:numcomplex} and we have used the fact that $J=\sqrt{J_1^2+J_2^2}$ is drawn from a Rayleigh distribution when $J_1, J_2$ are Gaussian-distributed. We see that the disorder average works beautifully in producing a complexity plateau, and that the complexity continues to grow linearly at early times, albeit with a modified slope $\sqrt{\frac{\pi}{2}} \sigma$. Heuristically, we see that even on as simple a manifold as $S^3$, the disorder average causes a kind of interference between many different random samples of the couplings $J_1, J_2$. In terms of the triangle waves seen in Fig.~\ref{fig:numcomplex}, if one imagines a large number of copies of the system with different values of $J$, after the copy with the maximum value of $J$ hits the first peak the various triangle waves begin to interfere destructively. At any fixed late time, the height of any one copy of the system is uniformly distributed between $0$ and $\pi$ and therefore the average complexity at long times is $\pi/2$.\footnote{We thank Cl\'{e}lia de Mulatier for discussion on this point.} It is straightforward to prove the late-time limit using the Fourier expansion of the triangle wave. Namely, the Fourier expansion of the triangle wave with slope $J$ can be written
\begin{align}
    f_{\triangle} (t) = \frac{\pi}{2} - \frac{4}{\pi} \sum_{n=1}^{\infty} \frac{1}{(2n-1)^2} \cos ((2n-1) Jt)
\end{align}
The disorder average can be performed term-by-term following \eqref{eq:averagedcomplex} assuming the integration can be exchanged with the infinite sum, and the result is
\begin{align}
    \bar{C}(t) = \frac{\pi}{2} - \frac{4}{\pi} \sum_{n=1}^{\infty} \frac{1}{(2n-1)^2} \left(1-\sqrt{2}\, t\sigma (2n-1) \mathcal{F} \bigl[  \frac{t \sigma}{\sqrt{2}} (2n-1) \bigr]\right) \label{eq:fouriercomplex}
\end{align}
where $\mathcal{F}[x] = e^{-x^2} \int_0^x e^{y^2} dy$ is Dawson's integral. If we allow ourselves to exchange the long-time limit with the infinite sum and use the identity $\lim_{x\to \infty} x \mathcal{F}[x] = \frac12$, one finds from \eqref{eq:fouriercomplex} that $\lim_{t \to \infty} \bar{C} (t) = \frac{\pi}{2}$. In the large-$N$ SYK model, we expect that the geometry of $SU(2^{N/2})$ is sufficiently involved that there is a ``self-averaging'' effect on complexity, namely that the late-time complexity saturation occurs in a single realization of the SYK model (without disorder averaging). This is further discussed in section \ref{sec:LTS}. 

It is also of interest to find the time scale for the onset of the plateau. Previous discussions of complexity in holography \cite{Brown2017} have noted that in large-$N$ chaotic systems this time scale ought to be exponential in the size of the system / number of qubits. The $N=2$ model studied in this section is neither large-$N$, nor chaotic; nevertheless it is useful to check that our results are compatible with an exponential time scale if $N$ were increased. Since the disorder-averaged complexity scales at early times like $\sigma t$, the plateau begins approximately upon reaching the asymptotic value at $t \sim \frac{\mathcal{C}_{\text{max}}}{\sigma}$. Assuming that $\mathcal{C}_{\text{max}} \sim \text{poly}(N) e^N$  and using the fact that in the SYK model, $1/\sigma$ is typically taken to scale polynomially in $N$, one finds a result $t \sim \text{poly}(N) e^N$ that is indeed consistent with the holographic expectation for the onset of the plateau.

 \begin{figure}[hbtp!]
\begin{center}
\includegraphics[width=.8\textwidth]{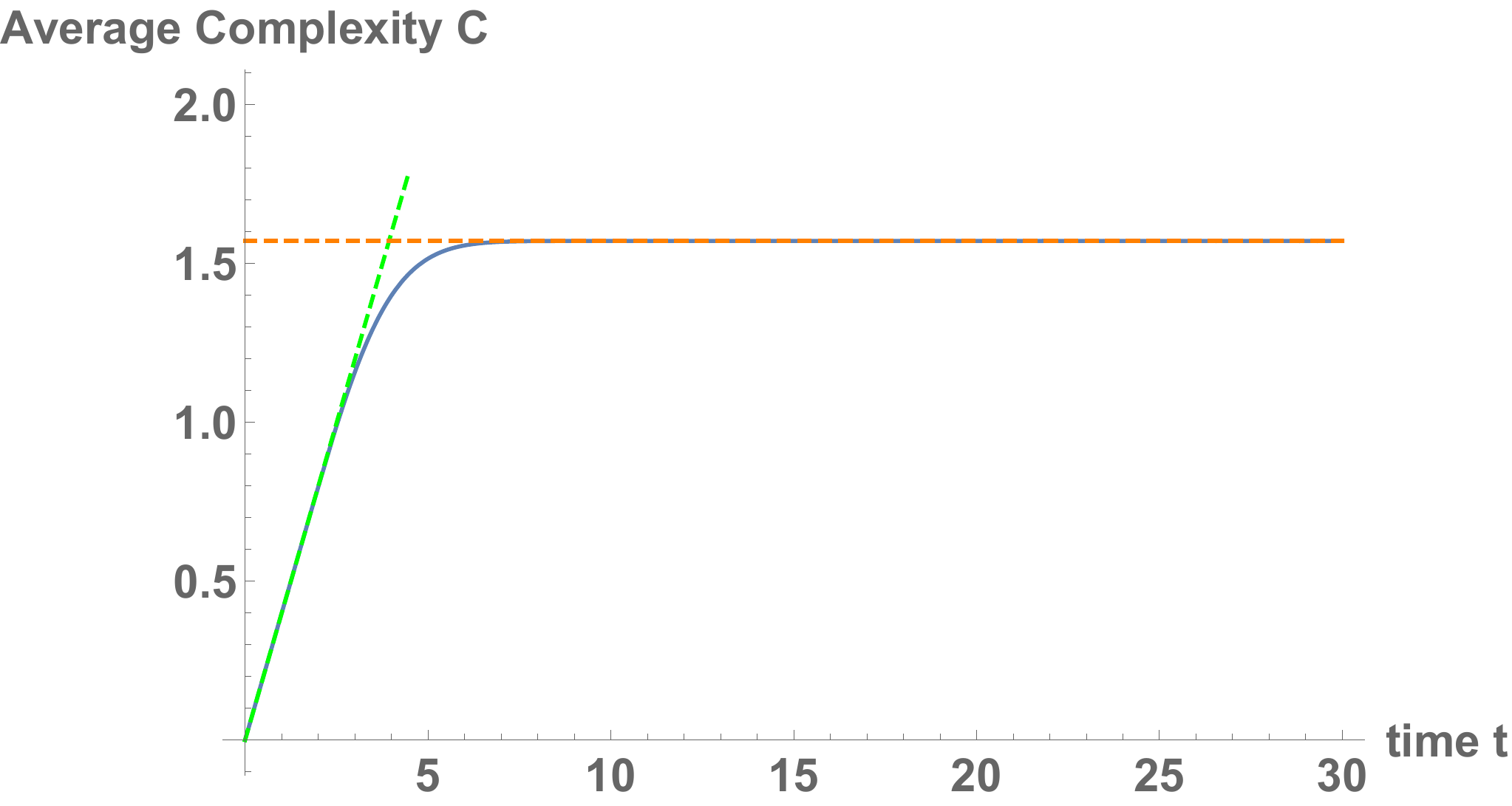} \caption{Disorder-averaged complexity (blue) choosing the magnitude $J$ of the couplings to be Gaussian-distributed rather than Rayleigh-distributed. The peak of Fig.~\ref{fig:complexaverage} is eliminated and the slope at early times is modified to $\sqrt{\frac{2}{\pi}} \sigma$ (green), though the plateau value of $\pi/2$ (orange) remains the same. \label{fig:smoothplateau}}
\end{center}
\end{figure}
Lastly, we remark that the existence of a peak in Fig.~\ref{fig:complexaverage} appears to be a peculiarity of taking $J_1, J_2$ to be Gaussian-distributed as our source of disorder; if one chose $J$ to be Gaussian-distributed rather than Rayleigh-distributed, the peak vanishes and the average complexity smoothly approaches a plateau, albeit with slope $\sqrt{\frac{2}{\pi}} \sigma$ (Fig.~\ref{fig:smoothplateau}).

\subsection{Linear geodesic for arbitrary $N$}
\label{sec:generalNlin}
For general $N$, the algebra $\mathfrak{su}(2^{N/2})$ is quite complicated.
We have collected some facts, including a derivation of the structure constants, in appendix \ref{sec:basis}.
However, the most important points for us are the following:
firstly,  the structure constants in the basis $T_i$ (corresponding to ordered products of gamma matrices) are fully antisymmetric by virtue of orthogonality in the trace norm. Secondly, ${f_{ij}}^\ell$ is nonzero if and only if 
\begin{equation}
{f_{ij}}^\ell \neq 0 \Leftrightarrow  i \oplus j = \ell,\ q_i q_j + q_{i\land j} \equiv 1 \mod 2.
\end{equation}
Here we are thinking of the multi-indices $i,j\cdots$ as binary numbers; for instance (in the ordering convention of appendix \ref{sec:basis}), the operator $T_3 = i \gamma_1\gamma_2$ corresponds to the binary number $00\cdots 0011$, $T_5 = i \gamma_1\gamma_3$ corresponds to the binary number $00\cdots 0101$ etc. Further, $q_i$ is the number of ones (i.e., the number of fermions) in $i$, $\oplus$ stands for the bitwise XOR and $\wedge$ stands for bitwise AND. The (suitably normalized) Cartan-Killing form follows after a short computation\footnote{The appropriate normalization factor in the general case with our choice of generators is $2^{-(N+1)}$.}
\begin{equation}
K_{ij} = \delta_{ij} .
\end{equation}
For convenience, we will label operators with $k$ or fewer fermions with undotted Greek indices $\alpha,\beta...$ and those with more than $k$ fermions with dotted Greek indices $\dot\alpha,\dot\beta...$, where $k < N$ is arbitrary for now. We choose the easy directions, i.e., operators with $k$ or fewer fermions, to have cost factors $c_\alpha = c$ and the hard directions, i.e., operators with more than $k$ fermions, to have cost factors $c_{\dot{\alpha}} = \bar{c}$. The Euler-Arnold equation can then be written schematically (since we have not determined overall sign)
\begin{equation}
c_i \frac{dV^i}{ds} = 2 (\bar{c}-c) \sum_{\substack{j,\ell\,\text{s.t}\, q_j q_\ell + q_{j \land \ell} \in 2\mathbb{Z}+1 \\ j \oplus \ell = i}} \pm V^j V^\ell ,
\label{eq:largeNeulerarnold}
\end{equation}
where we have explicitly written out the sums, and the index $i$ on the left-hand side is not to be summed over. There is an interesting structure to \eqref{eq:largeNeulerarnold} that emerges when we split into local and nonlocal directions.
We first observe that ${f_{\alpha\dot{\alpha}}}^\beta = -{f_{\beta\dot{\alpha}}}^\alpha$.
So, if a nonlocal direction with index $\dot{\alpha}$ appears in a local direction $\beta$'s velocity equation, it also appears in the velocity equation of the local direction $\alpha$ which multiplies it in $\beta$'s equation.
A similar story occurs for ${f_{\dot{\alpha}\alpha}}^{\dot{\beta}} = -{f_{\dot{\beta}\alpha}}^{\dot{\alpha}}$ for the nonlocal directions $\dot{\beta}$ and $\dot{\alpha}$.
The local direction $\alpha$ will occur in both of their velocity equations, appearing with opposite sign.
We can introduce antisymmetric matrices ${\dot{M}^{\alpha}}_\beta$ and ${M^{\dot{\alpha}}}_{\dot{\beta}}$ and rewrite \eqref{eq:largeNeulerarnold} as
\begin{equation}
\begin{split}
c \frac{dV^\alpha}{ds} & = 2(\bar{c}-c) {\dot{M}^{\alpha}}_\beta (V^{\dot{\gamma}}) V^\beta \\
\bar{c} \frac{dV^{\dot{\alpha}}}{ds} & = 2(\bar{c}-c) {M^{\dot{\alpha}}}_{\dot{\beta}} (V^\gamma) V^{\dot{\beta}} ,
\end{split}
\label{eq:spliteulerarnold}
\end{equation}
where $\dot{M}(V)$ is a matrix with {\it local} indices which depends linearly on the {\it nonlocal} directions' velocities and $M(V)$ is a matrix with {\it nonlocal} indices that depends linearly on the {\it local} directions' velocities.
Though this system is tricky to even write at arbitrary $N$, we can find a simple solution to it \emph{within the local subspace} using the ansatz:
\begin{equation}
\begin{split}
V^\alpha(s) & = v^\alpha ,\\
V^{\dot{\alpha}}(s) & = 0 ,
\end{split}
\label{eq:Nvelocities}
\end{equation}
which solves \eqref{eq:spliteulerarnold} because $\dot{M} = 0$.
The complexity is then
\begin{equation}
\mathcal{C} = \sqrt{\sum_\alpha (v^\alpha)^2} ,
\label{eq:Ncomplexity}
\end{equation}
where we have no contribution from the nonlocal directions.
This is in accord with our intuitions about quantum circuit construction, where we do not just suppress nonlocal gates but completely disallow them.
Since the velocities in \eqref{eq:Nvelocities} are constant, the path-ordering in \eqref{eq:path-ordered-solution} is trivial and the unitary path is
\begin{equation}
U(s) = e^{i v^\alpha T_\alpha s}.
\label{eq:Npathsolution}
\end{equation}
If we take our target state to be $U_{\text{target}} = e^{iHt}$ where $H$ is a $k$-local Hamiltonian
\begin{equation}
H = J^\alpha T_\alpha ,
\label{eq:Nhamiltonian}
\end{equation}
we can solve the boundary condition \eqref{eq:boundarycondition} to find one easy solution\footnote{One might think that the ambiguity in the logarithm gives multiple solutions here, but this is not the case, because generically the ``other solutions'' obtained from the log will not be entirely along the easy directions, and so are not admissible.}
\begin{equation}
v^\alpha = J^\alpha t.
\label{eq:Ntrivialsolution}
\end{equation}
We will refer to this geodesic as the \emph{linear geodesic}. Assuming that the linear geodesic is the correct minimum, we find that the complexity \eqref{eq:Ncomplexity} is
\begin{equation}
\mathcal{C} = t\sqrt{\sum_\alpha (J^\alpha)^2} = t\sqrt{e^{-S}\sum_{m=1}^{e^S}E_m^2},
\label{eq:Nlineargrowth}
\end{equation}
where in the second equality we have rewritten the coefficient in terms of the energy eigenvalues $E_m$ by relating the expressions inside the square roots to $\tr H^2$. The linear growth of the complexity in \eqref{eq:Nlineargrowth} matches expectations from holographic calculations of complexity as well as old observations about complexity growth in the geodesic formalism \cite{Nielsen2006,Brown2017}. Our task now is to investigate the validity of the assumption that the linear geodesic is the correct minimum to consider.

\begin{section}{Conjugate points and the eigenstate complexity hypothesis}\label{sec:conjugate}
One might wonder where the late-time behavior (i.e., late-time saturation) of complexity is going to appear from the previous discussion. The point, of course, is that the linear geodesic cannot be minimal for all times.
After all, $SU(2^{N/2})$ is a compact manifold, and no geodesic path on a compact manifold can globally minimize the length between the identity and $U = e^{-iHt}$ for all $t$.
In general, there are two ways a geodesic can become non-minimizing in a Riemannian manifold $M$:

\begin{enumerate}[1.]
    \item 
    \textbf{Conjugate points}: given a geodesic $U(s): [0,1] \to M$, there exists a variation through curves $U(\eta , s): [-\delta, \delta] \times [0,1] \to M$ such that $U(\eta, s)$ obeys the geodesic equation at first order in $\eta$, $U(0,s) = U(s)$, $U(\eta,0) = 1$ and $U(\eta , 1) = U(1) + \mathcal{O}(\eta^2)$.\footnote{See \cite{Witten:2019qhl} for a recent discussion of conjugate points in general relativity.}
    
    \item
    \textbf{Geodesic loops}: given a geodesic $U(s) : [0,1] \to M$ there is another geodesic $\widetilde{U}(s): [0,1] \to M$ such that $U$ and $\widetilde{U}$ have the same length $L[U] = L[\widetilde{U}]$, $U(0) = \widetilde{U}(0)$, and $U(1) = \widetilde{U}(1)$.
\end{enumerate}
These two conditions can roughly be thought of as local and global obstructions to minimality, respectively. This is because conjugate points along a geodesic segment mean that the segment is a saddle point,\footnote{We mean here a saddle point of the energy functional on the space of paths \cite{Morse1934}.} not a minimum; the number of conjugate points along the segment is equal to the number of ``downward directions''. Therefore, conjugate points are an obstruction to a geodesic segment being locally minimizing. On the other hand, the absence of conjugate points but presence of geodesic loops indicates that the geodesic segment is locally minimizing but \emph{not} globally minimizing. We will address the issue of conjugate points in this section. We will not prove the nonexistence of geodesic loops, but see the Discussion (section \ref{sec:discussion}) for some further comments.

Prior studies of complexity using toy models have largely avoided the question of conjugate points (although see \cite{Nielsen2007}, where the importance of conjugate points in circuit complexity was emphasized previously)  roughly by assuming all sectional curvatures are negative, so that geodesics originating at the same point generically diverge \cite{Brown2017,Lin2018}.
However, this assumption is worrisome, because it is well-known that any unimodular Lie group with left- or right-invariant metric must possess some strictly positive sectional curvature, or else be completely flat \cite{Milnor1976}.
If the sectional curvature cannot be everywhere bounded above by zero, one cannot rule out the existence of conjugate points on general grounds.
Therefore, it is crucial to understand conjugate points on the full group manifold in the complexity metric \eqref{eq:groupmetric}.
Here we will show a lower bound on the distance from the origin to the first conjugate point along the linear geodesic $V = Ht$.
We will call the time at which the linear geodesic develops this first conjugate point $t_c$.

In order to find conjugate points, we look for a velocity perturbation $\delta V(s)$, also called a \emph{Jacobi field},\footnote{More precisely, the Jacobi field is the first order deformation of the original geodesic, and $\delta V(s)$ is its derivative pulled back to the identity. We will sometimes loosely refer to $\delta V$ itself as the Jacobi field.} which obeys a first order differential equation known as the Jacobi equation, with particular boundary conditions which we will state precisely later.
In section \ref{sec:conjugate-jacobi}, we solve the Jacobi equation for the velocity perturbation $\delta V(s)$.
In section \ref{sec:conjugate-target}, we compute the first order change $\delta U$ in the target unitary due to a velocity perturbation which obeys the Jacobi equation.
Setting this to zero gives a boundary condition for the Jacobi equation, which corresponds to having a conjugate point. In section \ref{sec:Biinv}, we will show that with the bi-invariant choice of metric (i.e., with the same cost factors for all generators in the Lie algebra), the linear geodesic has a large number of conjugate points. 
In particular the first conjugate point appears at $t_c = \frac{2\pi}{E_{\text{max}}- E_{\text{min}}}$, where $E_{\text{max/min}}$ are the largest and smallest eigenvalues of the Hamiltonian respectively. In section \ref{sec:conjugate-heavylight}, we will then return to the right-invariant case with a large cost factor for the hard directions in the set of generators. We will argue that at large-$N$ and for Hamiltonians which satisfy what we will call the \emph{eigenstate complexity hypothesis} (ECH), the linear geodesic segment from the identity to $e^{-iHt}$ does not have any conjugate points for sub-exponential times, and thus the linear geodesic is at least locally minimizing until times exponential in $N$. 

\subsection{Solving the Jacobi equation}\label{sec:conjugate-jacobi}
In order to discover a conjugate point, we must deform the base geodesic with a velocity perturbation $\delta V(s)$ which solves the geodesic equation to first order; this first order equation for $\delta V(s)$ is called the Jacobi equation.
The Jacobi equation in our context is obtained by studying the first order correction to the Euler-Arnold equation (around the original, unperturbed geodesic $V = Ht$) under a velocity perturbation $Ht \to Ht + \delta V(s)$ (see Fig.~\ref{fig:ConjugatePoint} for an illustration).
We will confine our attention to the case with all the easy cost factors being $c = 1$ and all the hard cost factors being $\bar{c} = 1 + \mu$ (where $\mu \sim e^S$ as explained previously), but it would be interesting to generalize our analysis to the case where the cost factors along the hard directions vary with the scale of non-locality. 
We can then write the Jacobi equation as
\begin{equation}
\begin{split}
i \frac{d \delta V_L(s)}{ds} & = \mu t [H,\delta V_{NL}(s)]_L \\
i \frac{d \delta V_{NL}(s)}{ds} & = \frac{\mu t}{1+\mu} [H, \delta V_{NL}(s)]_{NL}
\end{split}
\label{eq:jacobi}
\end{equation}
where the subscripts $L$ (local) and $NL$ (nonlocal) denote projection into the local and nonlocal subspaces, i.e. 
\begin{equation}
\begin{split}
    \delta V_L &= \frac{1}{2^{N/2}}\sum_{\alpha} \tr (\delta V T_\alpha ) T_\alpha \\
    \delta V_{NL} &= \frac{1}{2^{N/2}}\sum_{\dot{\alpha}} \tr (\delta V T_{\dot{\alpha}} ) T_{\dot{\alpha}}
\end{split}
\end{equation} 
where $T_{\alpha}$, $T_{\dot{\alpha}}$ are bases for the local and nonlocal subspaces respectively.

\begin{figure}[!h]
\centering
\includegraphics[height=4cm]{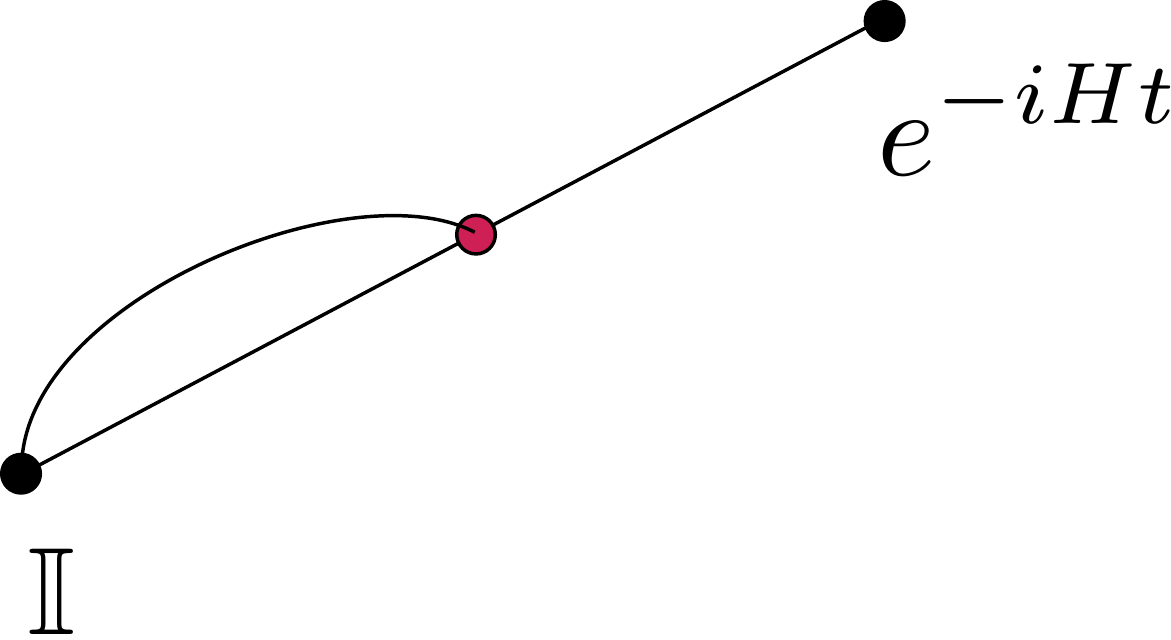}
\caption{\small{An illustration of a conjugate point, shown as the red point.}\label{fig:ConjugatePoint}}
\end{figure}
Let $\mathcal{A}_L$ denote the vector space spanned by the local generators in the Lie algebra, and $\mathcal{A}_{NL}$ denote the vector space spanned by the non-local generators. In order to solve the Jacobi equation, note that the second equation in \eqref{eq:jacobi} involves the super-operator $\mathbf{C}: \mathcal{A}_{NL} \to \mathcal{A}_{NL}$ defined by
\begin{equation}
\mathbf{C}(X) = \left[H, X\right]_{NL}.    
\end{equation}
The nonlocal equations can be solved by introducing a new basis $\widetilde{T}_{\dot{\alpha}}$ for the nonlocal subspace such that $\mathbf{C}$ is diagonal\footnote{This is always possible since $\mathbf{C}$ is Hermitian viewed as a matrix acting on $\mathcal{A}_{NL}$, and the spectral theorem of linear algebra states that Hermitian matrices may always be unitarily diagonalized with real eigenvalues.}:
\begin{equation}
\mathbf{C}(\widetilde{T}_{\dot\alpha}) = [H,\widetilde{T}_{\dot{\alpha}}]_{NL} = \lambda_{\dot{\alpha}} \widetilde{T}_{\dot{\alpha}} .
\end{equation}
In this basis, we can write $\delta V_{NL} = \sum_{\dot{\alpha}}\delta\widetilde{ V}^{\dot{\alpha}} \widetilde{T}_{\dot{\alpha}}$, where we note that the $\delta \widetilde{V}^{\dot\alpha}$ are numbers (i.e., the coefficients) while $\widetilde{T}_{\dot{\alpha}}$ are hard/non-local operators in the Lie algebra. The nonlocal equations become (in components)
\begin{equation}
i \frac{d \delta \widetilde{V}^{\dot{\alpha}}}{ds} = \frac{\mu t}{1+\mu} \lambda_{\dot{\alpha}} \delta \widetilde{V}^{\dot{\alpha}},
\end{equation}
with no summation over $\dot\alpha$ on the right-hand side. The solution is therefore
\begin{equation}
\delta \widetilde{V}^{\dot{\alpha}}(s) = \exp \left( \frac{-i\mu t \lambda_{\dot{\alpha}}s}{1+\mu} \right) \delta \widetilde{V}^{\dot{\alpha}}(0).
\label{eq:jacobi-nonlocal}
\end{equation}
Plugging this into the local equations, we have
\begin{equation}
i \frac{d \delta V_L}{ds} = \mu t \sum_{\dot{\alpha}} \exp \left( \frac{-i \mu t \lambda_{\dot{\alpha}}s}{1+\mu} \right) \delta \widetilde{V}^{\dot{\alpha}}(0) [H, \tilde{T}_{\dot{\alpha}}]_L .
\end{equation}
The local solution is
\begin{equation}
\delta V_L(s) = \delta V_L(0) - i\mu t \sum_{\dot{\alpha}} \frac{\exp \left( \frac{-i \mu t \lambda_{\dot{\alpha}}s}{1+\mu} \right) - 1}{\frac{-i \mu t \lambda_{\dot{\alpha}}}{1+\mu}} \delta \widetilde{V}^{\dot{\alpha}}(0) [H,\tilde{T}_{\dot{\alpha}}]_L .
\label{eq:jacobi-local}
\end{equation}

\subsection{Conjugate points as zero modes}\label{sec:conjugate-target}
We wish to use \eqref{eq:jacobi-nonlocal} and \eqref{eq:jacobi-local} to determine whether there are conjugate points.
This can be done by understanding the first order perturbation to the final unitary $U(1)$ induced by $\delta V$.
The exact final unitary and first order perturbation are, recalling \eqref{eq:path-ordered-solution},
\begin{equation}
U(1) = \mathcal{P} \exp \left( - i\int_0^1 ds (Ht + \delta V(s)) \right) = e^{-iHt} - i \delta U(1) ,
\end{equation}
where the $\delta U$ term is obtained by expanding the path-ordering in a Dyson series and taking the first term,
\begin{equation}\label{finDef}
U^{-1} \delta U(1) = \int_0^1 ds\, e^{iHts} \delta V(s) e^{-iHts} .
\end{equation}
We now define a super-operator $\mathbf{Y}_{(\mu)}: \delta V(0) \to U^{-1}\delta U(1)$ which acts by
\begin{equation}
\begin{split}
\mathbf{Y}_{(\mu)} ( \delta V(0)) & = \int_0^1 ds e^{iHts} \biggl[ \delta V_L(0) - i\mu t \sum_{\dot{\alpha}} \frac{\exp \left( \frac{-i \mu t \lambda_{\dot{\alpha}}s}{1+\mu} \right) - 1}{\frac{-i \mu t \lambda_{\dot{\alpha}}}{1+\mu}} \delta \widetilde{V}^{\dot{\alpha}}(0) [H,\tilde{T}_{\dot{\alpha}}]_L \\
& \hspace{.5cm} + \sum_{\dot{\alpha}} \exp \left( \frac{-i \mu t \lambda_{\dot{\alpha}}s}{1+\mu} \right) \delta \widetilde{V}^{\dot{\alpha}}(0) \tilde{T}_{\dot{\alpha}} \biggr] e^{-iHts},
\end{split}
\label{eq:superoperator}
\end{equation}
where we have inserted our solution for $\delta V(s)$ into equation \eqref{finDef}. A conjugate point, in this formalism, is given by the condition
\begin{equation}
    U^{-1}\delta U(1) = 0,
\end{equation}
and therefore corresponds to a zero mode of the super-operator \eqref{eq:superoperator}. So, our approach to finding conjugate points will be to study the spectrum of $\mathbf{Y}_{(\mu)}$ and check for when it develops zero modes. While this super-operator, as it appears in our analysis, is a linear operator on the Lie-algebra $\mathfrak{su}(2^{N/2})$, it is convenient to view $\mathbf{Y}_{(\mu)}$ as acting on the complexification of this vector space, i.e., on $\mathfrak{sl}(2^{N/2},\mathbb{C})$, and study the spectrum in this complexified space. The reason for doing this is that our Lie algebra is a vector space over a non-algebraically-closed field $\mathbb{R}$, and so the eigenvalues of $\mathbf{Y}_{(\mu)}$ need not be real, and the eigenvectors need not be real combinations of the elements of the Lie algebra. (This is true for essentially the same reason that solutions to $x^2 + 1 = 0$ only exist in $\mathbb{C}$ even though the equation involves coefficients only in $\mathbb{R}$.) Of course, in order for a true conjugate point to appear for some values of $t$ and $\mu$, the zero mode must be Hermitian and traceless. In other words, it must indeed be a valid element of $\mathfrak{su}(2^{N/2})$.

\subsection{Conjugate points in the bi-invariant case}
\label{sec:Biinv}
Solving for the conjugate points at general values of $\mu$ is analytically hard. We will only be able to do it approximately in section \ref{sec:conjugate-heavylight} for large-$N$ Hamiltonians which satisfy a certain complexity criterion on their eigenstates. But before doing that, it is useful to look at the much simpler case of $\mu = 0$ where we can obtain all the conjugate points exactly. This is because at $\mu = 0$, where all generators are considered computationally ``easy'', equation \eqref{eq:superoperator} simplifies greatly, and we get
\begin{equation}
    \mathbf{Y}_{(0)}(\delta V(0)) = \int_0^1 ds\,e^{iHts}\delta V(0) e^{-iHts}.
\end{equation}
It is easy to guess the eigenvectors of this super-operator: they are simply the energy eigenstate projectors $|m\rangle \langle n|$, where $|m\rangle$ and $|n\rangle$ are eigenstates of the Hamiltonian with energies $E_m$ and $E_n$ respectively. Indeed, we find
\begin{equation}
\mathbf{Y}_{(0)}(|m\rangle \langle n|) = \int_0^1 ds\,e^{i(E_m - E_n)ts}|m\rangle \langle n| = \phi_{mn}(t) |m\rangle \langle n|,
\end{equation}
where the eigenvalue $\phi_{mn}(t)$ is given by\footnote{Notice that diagonal projectors $\ket{n}\bra{n}$ have constant eigenvalue 1 and therefore cannot lead to conjugate points in the bi-invariant case.} 
\begin{equation}
  \phi_{mn}(t) \equiv \frac{e^{i\Delta_{mn}t} -1}{i\Delta_{mn}t} \;\;  ,\;\;\Delta_{mn}= (E_m - E_n).
  \label{eq:phi-function}
\end{equation}
The eigenvalue $\phi_{mn}$ (for $E_m\neq E_n$) becomes zero at 
\begin{equation}\label{biinvCPs}
    t_{mn} = \frac{2\pi}{\Delta_{mn}}\mathbb{Z}.
\end{equation}
Indeed, at these times, the eigenvalues corresponding to both $|m\rangle\langle n|$ and $|n\rangle\langle m|$ become zero, and we can construct two Hermitian linear combinations out of these. Therefore, the linear geodesic develops a large number of conjugate points at the times given by equation \eqref{biinvCPs}, for all the possible choices of $E_m$ and $E_n$. The first time $t>0$ at which it develops a conjugate point is 
\begin{equation}
    t_{c} = \frac{2\pi}{E_{\text{max}}- E_{\text{min}}} \equiv \frac{2\pi}{\Delta_{\text{max}}}.
\end{equation}

In the SYK model, the maximum separation is known to be $\Delta_{\text{max}} \sim N$, and so the linear geodesic stops being minimal after $t_c\sim \frac{2\pi}{N}$. However, this model is expected to be chaotic, so how is it that the conjugate points are appearing at a time of $O(1/N)$?
The resolution of course lies in the fact that the bi-invariant metric is not the correct Riemannian metric for complexity.  To understand physically relevant conjugate points we need to select a notion of locality for our generators.
In other words, we need to choose which operators in the theory are ``simple".
By choosing the bi-invariant metric on the generators  we have allowed arbitrary operators as local, but this is definitely not a physically sensible choice. However, the above calculation emphasizes the importance of conjugate points, and the need to make sure that they are absent if we are to establish the minimality of a geodesic.
We now turn to the question of what happens to conjugate points for chaotic systems when a suitable notion of locality has been established by turning on cost factors in the complexity metric.

\subsection{Turning on cost factors}\label{sec:conjugate-heavylight}
We will turn on a finite cost factor $\mu$, which will separate ``easy'' and ``hard'' computational directions, or, more physically, operations that we will consider ``local'' or ``non-local''. Our aim is to show that the linear geodesic is locally minimizing for times exponential in $N$, and so contains no conjugate points till such time. As stated previously, we do not have an exact solution for the spectrum of $\mathbf{Y}_{(\mu)}$ (although it is possible to calculate this spectrum perturbatively in $\mu$, see appendix \ref{app:ConjugatePoints}). However, if the Hamiltonian is sufficiently chaotic, then the situation simplifies greatly. More precisely, if the off-diagonal eigenstate projectors $|m\rangle \langle n|$ of the Hamiltonian are ``complex'', in the sense that their overlaps with the local generators are exponentially suppressed in $N$, then we can give an approximate formula for the spectrum of the super-operator $\mathbf{Y}_{(\mu)}$ at finite $\mu$. We will call this criterion the eigenstate complexity hypothesis, or ECH for short:

\noindent\textbf{Eigenstate Complexity Hypothesis (ECH)}: Let $H$ be the Hamiltonian with energy eigenstates $|m\rangle, |n\rangle$ etc., $T_{\alpha}$ be the local generators in the Lie algebra, and $T_{\dot\alpha}$ be the non-local generators. Define
\begin{equation}\label{RDef}
    R_{mn} = \frac{\sum_{\alpha} |\langle m|T_{\alpha}|n\rangle |^2}{\sum_{\alpha} |\langle m|T_{\alpha}|n\rangle |^2+\sum_{\dot\alpha} |\langle m|T_{\dot\alpha}|n\rangle |^2}.
\end{equation}
We will say that the Hamiltonian and the gate set satisfy the eigenstate complexity hypothesis, if in the large-$N$ limit for $E_m\neq E_n$,
\begin{equation}
    R_{mn} = e^{-2S} \text{poly}(S)\, r_{mn},\label{eqECH2}
\end{equation}
where $S$ is $\ln \text{dim}$ of the Hilbert space (i.e., $S = \frac{N}{2}\ln\,2$ for the SYK model), $\text{poly}(S)$ is some polynomial in $S$, and $r_{mn}$ are $O(1)$ numbers which do not scale with $N$. We can equivalently state this as 
\begin{equation}
||\,\ket{m}\bra{n}_L\,|| = O(e^{- S} \text{poly}(S)),
\label{eq:no-local-approx}
\end{equation}
where recall that the subscript $L$ indicates projection to the local/easy subspace in the Lie algebra and the operator norm is defined by $||X|| = [\mathrm{Tr}(X^{\dagger}X)]^{1/2}$.

The physical intuition behind this criterion is that off-diagonal projectors of the form $\ket{m}\bra{n}$ map the energy eigenstate $\ket{n}$ to a different eigenstate $\ket{m}$. For chaotic Hamiltonians, this operation should be complex from the point of view of local generators in the Lie algebra, since we expect these energy eigenstates to differ in their fine-grained microstructure. Another reason to expect ECH is that for sufficiently chaotic Hamiltonians, off-diagonal projectors like $|m\rangle \langle n|$ tend to have a uniformly distributed overlap with the generators in the Lie algebra (see Fig.~\ref{cDist} in appendix \ref{app:ECH}), and since there are exponentially many non-local generators and only polynomially many local generators (assuming $k$ does not scale with $N$), the projection of $|m\rangle \langle n|$ onto the local directions should be exponentially suppressed in $N$, as per equation \eqref{eqECH2}.

\begin{figure}[t]
\centering
\begin{tabular}{c c c}
   \includegraphics[height=5cm]{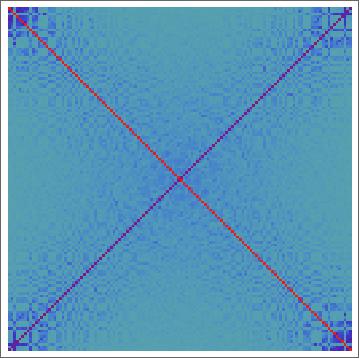}  &  \includegraphics[height=4.5cm]{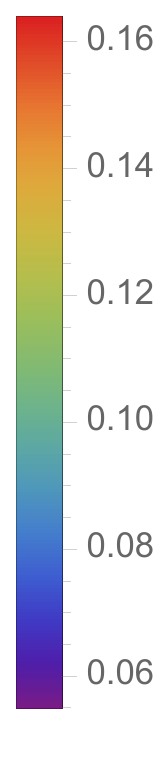}& \includegraphics[height=5cm]{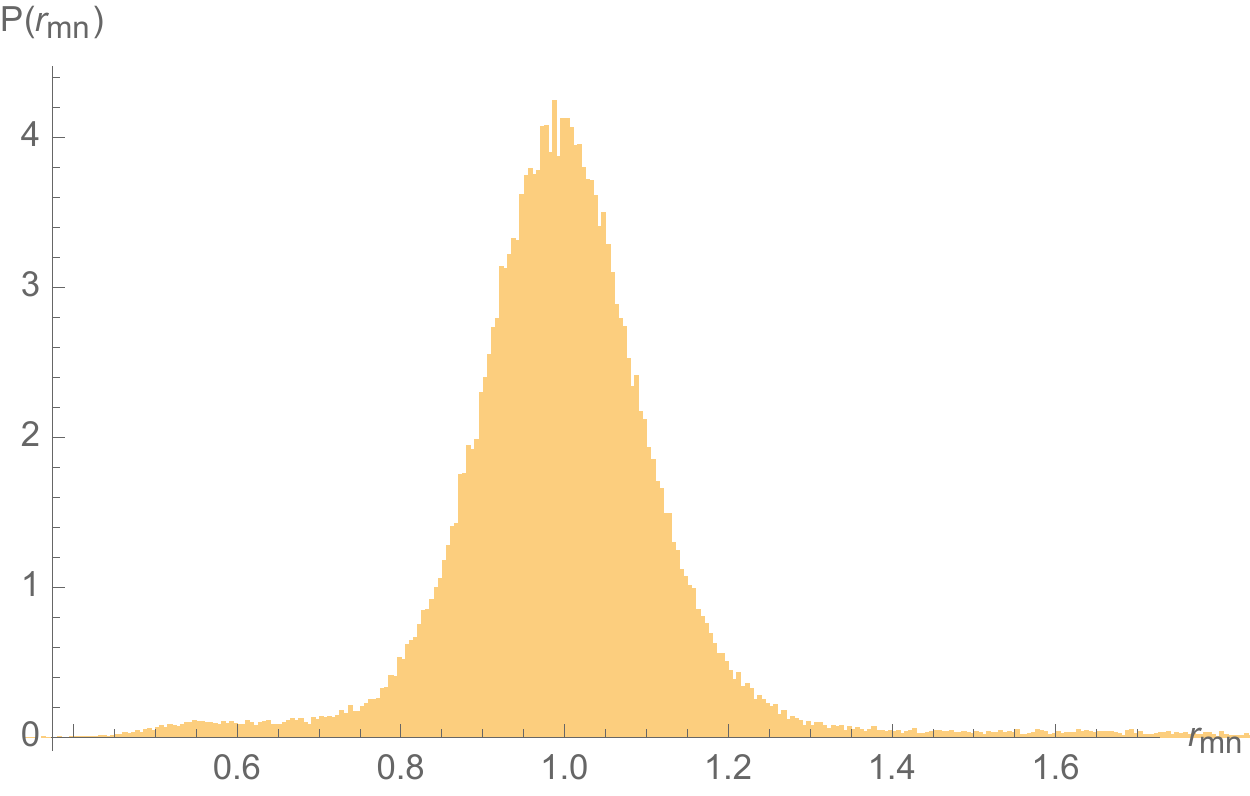}
\end{tabular}
\caption{\small{(Left) A visualization of the matrix $R_{mn}$ for the SYK model at $N=14, k=4, q=3, \mathcal{J}=1$, for a single realization. Note that off-diagonal elements are suppressed. The vast majority of the off-diagonal matrix elements are close to $\sim 0.09$, which is precisely the number of local generators divided by the total number of generators. The diagonal elements seem to be enhanced compared with the rest. (Right) A histogram of $r_{mn}$s defined in equation \eqref{rdef} for $N=12,k=3,q=3$ for 100 realizations. }\label{ECH}}
\end{figure}

The interacting SYK model satisfies the ECH. To demonstrate this, we have shown an array-plot of the matrix $R_{mn}$ for a single realization in the left panel of Fig.~\ref{ECH}, for the SYK model at $N= 14$, $\mathcal{J}=1$ and $k=4, q= 3$. We see that the off-diagonal elements of $R_{mn}$ are indeed exponentially suppressed. Taking 
\begin{equation}\label{rdef}
    R_{mn} = \frac{n_{\text{easy}}}{2^{N}}r_{mn},
\end{equation}
where the $N$-dependent coefficient is the number of easy generators divided by the total number of generators, we have shown the distribution $P(r_{mn})$ of all the $r_{mn}$s (including diagonals) over 100 realizations of the SYK model in the right panel of Fig.~\ref{ECH} (with $N=12$ for convenience). The $r_{mn}$s are distributed with a (sample) mean of $\bar{r}_s=1$, and (sample) standard deviation of $\sigma_s = 0.14$.\footnote{It is easy to show from the definition of $r_{mn}$ that their mean is one: $\frac{1}{2^N}\sum_{m,n} r_{mn} = 1$. The distribution $P(r_{mn})$ can be roughly approximated by the normal distribution with mean $\bar{r}\simeq 1$ and standard deviation $\sigma \simeq 0.098$. A slightly better approximation is provided by Student's t-distribution with the parameters $\bar{r}=0.994, \sigma= 0.093$ and the number of degrees of freedom $\nu= 6$.} We have also checked other values of $N$ and $(k,q)$ (with $q<k$) and found similar behavior. One novelty for $q$ even (see appendix \ref{app:ECH} for further discussion) is that the Hamiltonian has a fermion number symmetry (which additionally is diagonal in the basis involving products of fermions), and this leads to an $O(1)$ splitting of the distribution $P(r_{mn})$ into two distributions, corresponding to the off-diagonal projectors which either preserve or reverse the fermion number symmetry. 

So far, we have presented some numerical evidence to show that the SYK model satisfies the ECH. More generally, we expect chaotic Hamiltonians to satisfy ECH (provided an appropriate choice is made for the $k$-local generators) as a consequence of a form of the \emph{eigenstate thermalization hypothesis} (ETH), which is believed to be true in general chaotic quantum systems \cite{PhysRevA.43.2046, 1994PhRvE..50..888S, DAlessio:2016rwt} (see \cite{Sonner:2017hxc, Hunter-Jones:2017raw, Nayak:2019khe} for discussion of ETH in the SYK model). ECH is of course very reminiscent of the ETH. In fact, we can see how the two are related in the SYK model. If we take the generators to be $T_{i} \sim \psi_{a_1}\cdots \psi_{a_m}$, the denominator in the definition (equation \eqref{RDef}) of $R_{mn}$ is equal to $e^{S} = 2^{N/2}$; this just follows from the fact that $\ket{m}\bra{n}$ has operator norm one, while each of the generators $T_i$ has norm $e^{S/2}$. Now, if we further assume that the local/easy generators satisfy ETH, then each term in the numerator of $R_{mn}$ is also $O(e^{-S})$.\footnote{More precisely, the ETH suppression to $R_{mn}$ is $e^{-S(\bar{E})}$ where $\bar{E}=(E_m+E_n)/2$, but we expect $e^{S-S(\bar E)}$ to be polynomial in $S$.}  Since there are at most polynomially many local/easy generators (assuming $k$ does not scale with $N$), we deduce that $R_{mn} = O(\text{poly}(S) e^{-2S})$, provided the easy generators satisfy ETH. From this perspective, we may view ECH as saying that the easy generators in our choice of the gate set should satisfy ETH, but where our easy generators are $k$-local, and so involve multi-site operators (not simply 1-local operators). On the other hand, ECH is a logically independent criterion from ETH; it requires that the off-diagonal outer products $\ket{m}\bra{n}$ have small projection onto the easy/local directions, i.e., that they are \emph{complex}, or alternatively that they are uniformly distributed in terms of their overlaps with all the $e^{2S}$ generators of the gate-set. We expect that large-$N$ integrable Hamiltonians should violate ECH, but it would be interesting to study this in greater detail. Certainly, off-diagonal operators of the form $\ket{m}\bra{n}$ in integrable systems tend to have overlaps with a far smaller subset of the $e^{2S}$ generators in the gate set (see Fig.~\ref{fig:cDistInt} in appendix \ref{app:ECH}.) Since the norm of $\ket{m}\bra{n}$ is one, this naturally requires the individual overlaps $\langle n|T_i|m\rangle$ to be larger. For instance, in appendix \ref{app:ECH} we show numerical evidence that for $q=2$ (quadratic) SYK-like Hamiltonians, the individual overlaps $\langle n|T_i|m\rangle$ are all $O(1)$, and hence $R_{mn}$ only has an $e^{-S}$ suppression, as opposed to $e^{-2S}$ in the chaotic case (or equivalently $||\, |m\rangle\langle n|_L|| \sim e^{-S/2}$ as opposed to $e^{-S}$). Indeed, our arguments below for complexity growth will crucially rely on this enhanced suppression in chaotic systems.  

Let us now return to the problem of conjugate points at finite cost factor. If we take the statement \eqref{eq:no-local-approx} of ECH as given, then we have
\begin{eqnarray}
\mathbf{C}(\ket{m}\bra{n}_{NL}) &=& [H, \ket{m}\bra{n}_{NL}]_{NL}  \nonumber\\
&=& [H, \ket{m}\bra{n}]_{NL} - [H, \ket{m}\bra{n}_L]_{NL} \nonumber\\
&=& \Delta_{mn} \ket{m}\bra{n}- [H, \ket{m}\bra{n}_L]_{NL}-\Delta_{mn} \ket{m}\bra{n}_{L} \nonumber\\
&=& \Delta_{mn} \ket{m}\bra{n} + O(e^{-S} \text{poly}(S)),
\label{eq:C-approx}
\end{eqnarray}
where in the last line we have used ECH together with the fact that the Hamiltonian is a linear combination of only  polynomially many generators, and so the norm of  $\left[H,\ket{m}\bra{n}_L\right]_{NL}$ can at most get a polynomial enhancement over the exponentially suppressed norm of $\ket{m}\bra{n}_L$. This implies that if we take our initial velocity to be $\delta V(0) = \ket{m}\bra{n}$, then the solution \eqref{eq:jacobi-nonlocal}, \eqref{eq:jacobi-local} to the Jacobi equation simplifies substantially
\begin{equation}\label{Japprox1}
    \delta V_{mn}(s) = \exp\left(-\frac{i\mu t\Delta_{mn} s}{1+\mu}\right)\ket{m}\bra{n} + \cdots.
\end{equation}
Below, we will carefully justify that the corrections to equation \eqref{Japprox1}, denoted as $\cdots$ above, are exponentially suppressed in $N$, but for now we will proceed with the main argument. With equation \eqref{Japprox1} in hand, we can evaluate the action of the super-operator $\mathbf{Y}_{(\mu)}$ on $\ket{m}\bra{n}$:
\begin{eqnarray}
\mathbf{Y}_{(\mu)}(\ket{m}\bra{n}) &=& \int_0^1 ds\,e^{iHts}\delta V_{mn}(s)  e^{-iHts}  \nonumber\\
&=&  \int_0^1 ds\,\exp\left(\frac{i t\Delta_{mn} s}{1+\mu}\right)\ket{m}\bra{n} + \cdots,
\end{eqnarray}
where, once again, the correction terms are exponentially suppressed in $N$, as will be justified below. Performing the $s$ integration, we find
\begin{equation}
\mathbf{Y}_{(\mu)} (\ket{m}\bra{n}) = \frac{\exp \left( \frac{i \Delta_{mn} t}{1+\mu} \right) - 1}{\frac{i \Delta_{mn} t}{1+\mu}} \ket{m} \bra{n}+\cdots = \phi_{mn} \left( \frac{t}{1+\mu} \right) \ket{m} \bra{n}+ \cdots.
\label{eq:superop-eigenval-approx}
\end{equation}
(Note that the function $\phi_{mn}$ was defined in \eqref{eq:phi-function}.)
Therefore, under the assumption \eqref{eq:no-local-approx}, the super-operator $\mathbf{Y}_{(\mu)}$ is a diagonal matrix in the $\ket{m}\bra{n}$-basis with the diagonal entries given by $\phi_{mn}\left( \frac{t}{1+\mu} \right)$. The diagonal entry corresponding $\ket{m}\bra{n}$ becomes zero at 
\begin{equation}
    t_{mn}(\mu) \simeq \frac{2\pi(1+\mu)}{\Delta_{mn}}\mathbb{Z},
\end{equation}
up to $O(e^{-S}\text{poly}(S))$ corrections.\footnote{Each conjugate point is two-fold degenerate at leading order in $N$, and the exponentially suppressed corrections may split this two-fold degeneracy.} Therefore, as we crank up the cost factor $\mu$, all the diagonal entries become approximately equal to one. Indeed, the first time at which one of these diagonal entries becomes zero moves to later and later time (see Fig.~\ref{CPExp}) is now approximately located at
\begin{equation}
t_c \simeq \frac{2\pi}{\Delta_{\text{max}}} (1+\mu) .
\label{eq:conjugate-time}
\end{equation}
If we take the cost factor to be 
$$\mu \sim e^{(1-\epsilon) S},$$
where $\epsilon$ is some small positive number (as will become clear shortly, the above argument works when $\mu t \ll e^S$ which is satisfied by this choice at any sub-exponential time), then $\mathbf{Y}$ will be approximately the identity matrix for any time sub-exponential in $S$, and so we do not expect it to have zero modes. Of course, there is an important caveat here -- although the corrections to $\mathbf{Y}$ are exponentially small, the matrix in question is exponentially large and so one might worry that the eigenvalues of $\mathbf{Y}$ get corrected at $O(1)$. We will address this issue below. 

\begin{figure}[t]
\centering
\begin{tabular}{c c}
   \includegraphics[height=4.5cm]{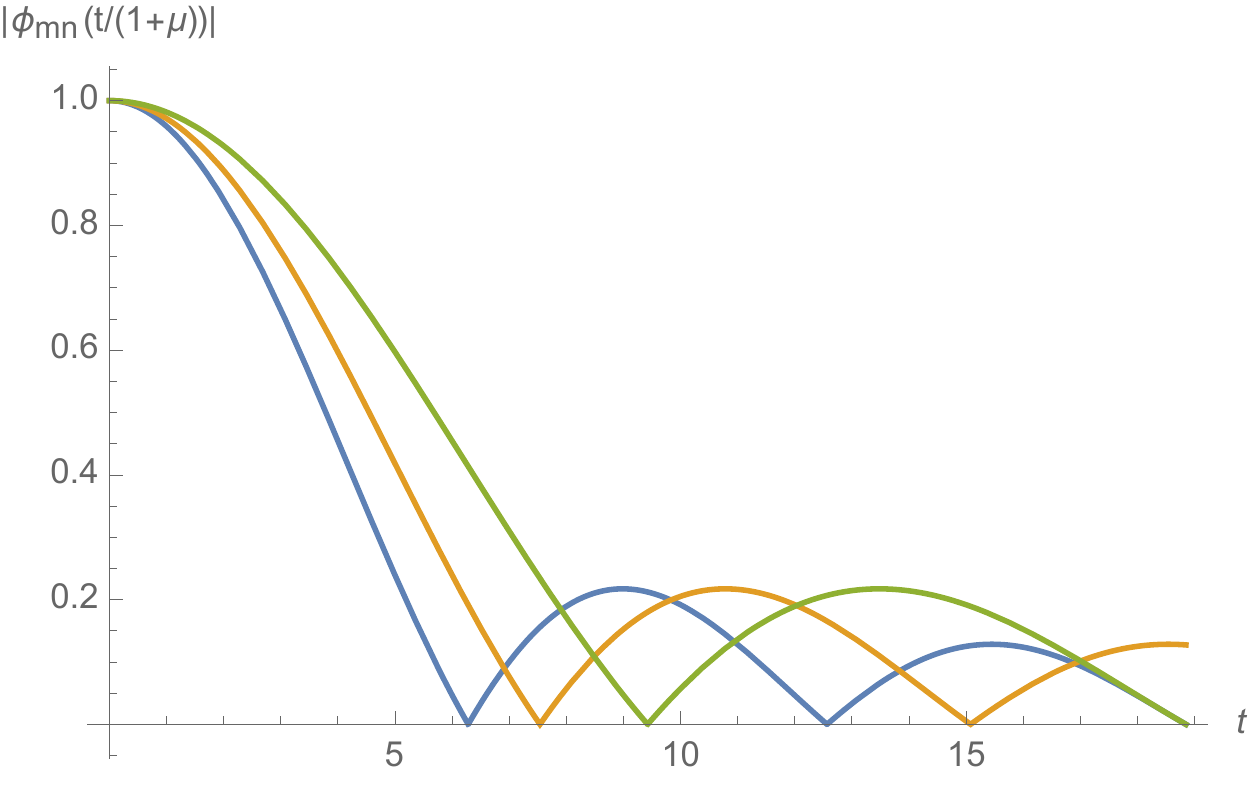}  &  \includegraphics[height=4.5cm]{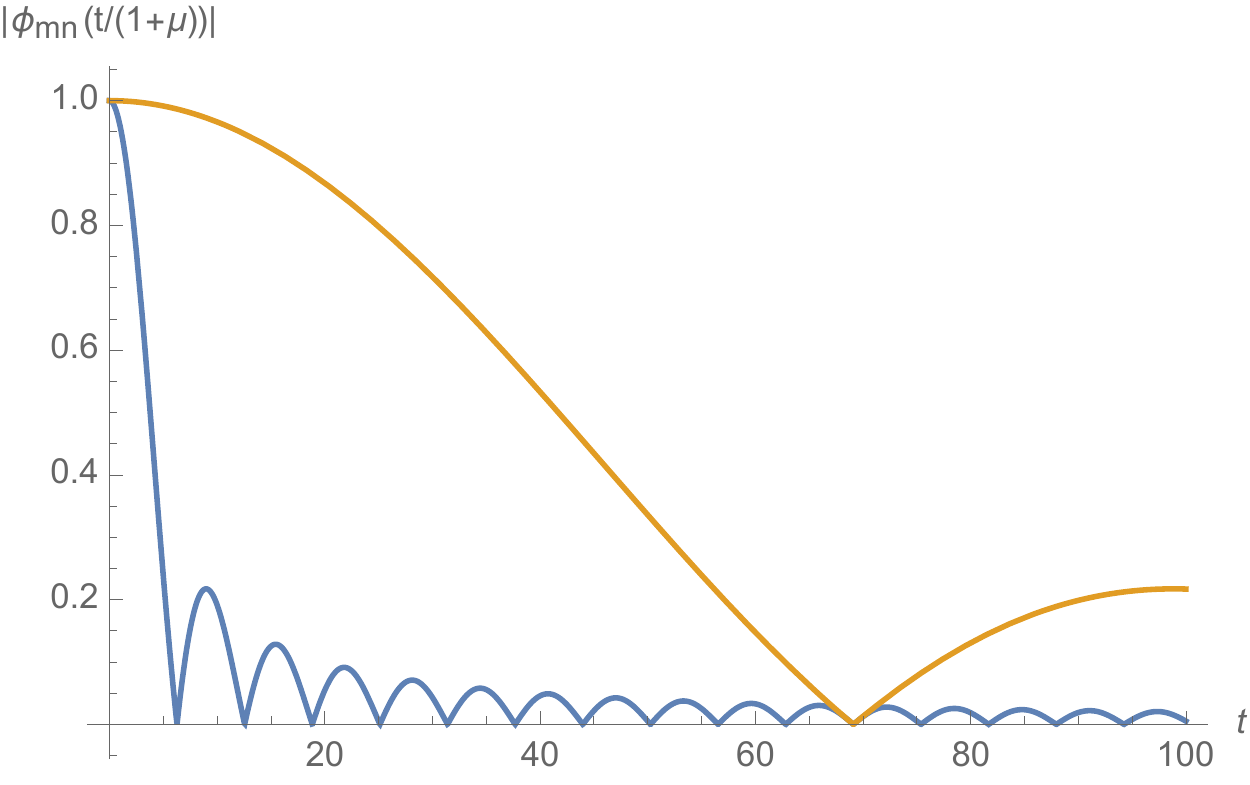}
\end{tabular}
\caption{\small{As we crank up $\mu$ from $\mu = 0$ (blue curves) to finite non-zero values, the conjugate points corresponding to $\ket{m}\bra{n}$ (zeros of $\phi_{mn}$) move towards larger times. The orange and green curves on the left correspond to $\mu=0.2$ and $\mu=0.5$ respectively, while the orange curve on the right corresponds to $\mu=10.$ We have taken $\Delta_{mn} =1$ for simplicity. }\label{CPExp}}
\end{figure}
Here we have assumed that $\Delta_{\text{max}}$ does not  scale exponentially with $N$. Indeed, for the SYK model $\Delta_{\text{max}}= O(N)$. This shows that the linear geodesic segment from the identity to $e^{-iHt}$ is locally minimizing for times exponential in $N$. To be precise, we have shown that all the low-lying conjugate points which were present in the bi-invariant case have moved to exponential time upon turning on the cost factor $\mu = e^{(1-\epsilon) S}$.  In the bi-invariant case, all the diagonal projectors $\ket{m}\bra{m}$ are eigenvectors of the super-operator $\mathbf{Y}_{(0)}$ with unit eigenvalue and do \emph{not} correspond to conjugate points.  We can argue from continuity that this is still the case when we turn on the cost factor $\mu$: since conjugate points are zero modes of $\mathbf{Y}_{(\mu)}$, they cannot simply appear out of nowhere; as we can see in Fig.~\ref{CPExp}, they can only move smoothly along the time axis. Therefore, no new conjugate points should appear at finite time with a finite cost factor.   This argument can be formalized using Morse theory \cite{Morse1934}. 

We also note here that if there is an off-diagonal projector $\ket{m}\bra{n}$ which violates ECH ``maximally'', namely that is has an almost unit overlap with the easy/local directions and a small overlap with the hard/non-local directions, then one can similarly show that such a projector corresponds to an approximate eigenvector of $\mathbf{Y}_{(\mu)}$ with the eigenvalue $\phi_{mn}(t)$. In this situation, we may expect to find conjugate points at the $O(1)$ times $t = \frac{2\pi}{\Delta_{mn}}\mathbb{Z}$, provided $\Delta_{mn}$ is not exponentially small, and if so the linear geodesic would stop being minimizing early on in time evolution. We expect this behavior to be present at small $N$.

\noindent\textbf{Bounding the correction terms}: Now we wish to carefully justify that all the correction terms which were ignored above are indeed exponentially suppressed. To this end, let $\delta V_{mn}(s)$ be the Jacobi field along the linear geodesic with the initial condition $\delta V_{mn}(0) = \ket{m}\bra{n}$, and define
\begin{equation}
    \delta V_{mn}(s) = c(s) \ket{m}\bra{n}+ \delta W(s), 
\end{equation}
where $c(s) = e^{-\frac{i\mu ts\Delta_{mn}}{1+\mu}}$, and $\delta W(s)$ is the correction to the leading order result in equation \eqref{Japprox1}. We insert this into the Jacobi equation to obtain the differential equations satisfied by $\delta W$:
\begin{equation}
\begin{split}
i \frac{d \delta W_L(s)}{ds} & = \mu t [H,\delta W_{NL}(s)]_L + \mathcal{S}_L(s),\\
i \frac{d \delta W_{NL}(s)}{ds} & = \frac{\mu t}{1+\mu} [H, \delta W_{NL}(s)]_{NL}+\mathcal{S}_{NL}(s),\\
\delta W(0) & = 0,
\end{split}
\label{Japprox2}
\end{equation}
where the source term $\mathcal{S}$ above is given by
\begin{equation}
\mathcal{S}(s) = \frac{\mu^2 t}{1+\mu}c(s)\Delta_{mn}\ket{m}\bra{n}_L -\mu t c(s)\left[H, \ket{m}\bra{n}_L\right]_L-\frac{\mu t}{1+\mu} c(s)\left[H, \ket{m}\bra{n}_L\right]_{NL}.
\end{equation}
As long as $\mu t \ll e^S$, say for instance $\mu t \sim e^{(1-\epsilon) S}$, then the source terms have an exponentially suppressed norm by ECH:
\begin{equation}
    ||\mathcal{S}|| = O(e^{- \epsilon S}\text{poly}(S)),
\end{equation}
where $S=\frac{N}{2}\ln(2)$, and we are using the Frobenius norm $||X||^2 = Tr(X^{\dagger}X)$. For polynomial times, we can therefore take $\mu \sim e^{(1-\epsilon)S}$, and the source terms will still be suppressed; beyond this value of $\mu$ our arguments here will break down.\footnote{Note that for integrable systems, weaker suppression implies that our argument breaks down at $\mu \sim e^{S/2}$, that is far before the required value of $e^S$ for the cost factor. For chaotic systems, we can push the cost factor to $e^{(1-\epsilon)S}$, which is almost the required value.} Expressing $\delta W$ in terms of the basis $(T_{\alpha}, \widetilde{T}_{\dot{\alpha}})$ introduced previously and solving the second equation in \eqref{Japprox2}, we obtain for the non-local piece of $\delta W$:
\begin{equation}
\delta W_{NL}(s) = -i\int_0^s ds'\sum_{\dot{\alpha}} e^{\frac{i\mu t(s-s')\lambda_{\dot{\alpha}}}{1+\mu}}\mathcal{S}_{\dot{\alpha}}\widetilde{T}_{\dot{\alpha}}.
\end{equation}
Therefore, the norm of this correction term is given by
\begin{equation}
|| \delta W_{NL}(s) || \leq \int_0^s ds' ||\mathcal{S}|| = O(e^{-\epsilon S}\text{poly}(S)),
\end{equation}
where we have used $|| \int X || \leq \int ||X||$. Repeating the same argument for the local directions, we see that in fact
\begin{equation}\label{Wbound}
|| \delta W(s) || = O(e^{- \epsilon S}\text{poly}(S)).
\end{equation}
Now coming to the action of the super-operator on $\ket{m}\bra{n}$, we have the exact statement
\begin{eqnarray}
    \mathbf{Y}_{(\mu)}(\ket{m}\bra{n}) &=& \int_0^1 ds\,e^{iHts}\delta V_{mn}(s)  e^{-iHts}  \nonumber\\
&=&  \phi_{mn}\left(\frac{t}{1+\mu}\right)\ket{m}\bra{n} + \int_0^1 ds\,e^{iHts}\delta W(s)  e^{-iHts}.
\end{eqnarray}
We can bound the norm of the second term above by once again using $|| \int X || \leq \int ||X||$, together with equation \eqref{Wbound}:
\begin{eqnarray}
||\int_0^1 ds\,e^{iHts}\delta W(s)  e^{-iHts}|| &\leq& \int_0^1 ds\,||e^{iHts}\delta W(s)  e^{-iHts}||\nonumber\\
&=& \int_0^1 ds\,||\delta W(s)|| \nonumber\\
&=& O(e^{- \epsilon S}\text{poly}(S)).
\label{eq:suppression}
\end{eqnarray}
This completes our justification that the corrections to equation \eqref{eq:superop-eigenval-approx} are indeed exponentially suppressed in $N$. 
The upshot of these arguments is that, for the parameter regimes we are interested in, the functions $\phi_{mn}\left( \frac{t}{1+\mu} \right)$ remain close to one and all other contributions are suppressed.
Therefore, no zero modes of the super-operator can develop before at least one $\phi_{mn}$ has dropped away from 1, and this does not occur until times exponential in $N$.

We will now address a possible caveat in the above discussion: we have shown that the superoperator $\textbf{Y}$ is an approximately diagonal matrix with the diagonal entries being approximately one at times much smaller than $t_c$, and exponentially suppressed off-diagonal entries. So let us write
\begin{equation}
    \textbf{Y} = \Phi + \delta\textbf{Y} ,
\end{equation}
where $\Phi$ is the diagonal part and $\delta\textbf{Y}$ is the off-diagonal part. Equation \eqref{eq:suppression} shows that the $L_2$ norm of any row in $\delta\textbf{Y}$ is bounded by an exponentially small quantity.
One might worry that, since there are exponentially many of these rows, they may combine to lead to significant deviations in  the eigenvalues of $\textbf{Y}$ compared to $\Phi$.\footnote{We thank Daniel Ranard for emphasizing this point to us.} The point, however, is that the constraint on the norm of the individual rows of $\delta\textbf{Y}$ is strong enough that almost all the exponentially many eigenvalues can only receive exponentially small corrections, while only an $O(1)$ number of eigenvalues can be affected significantly. We can see this by estimating the average magnitude of the eigenvalues of $\delta\textbf{Y}$,\footnote{Note that we can focus on the eigenvalues of $\delta\textbf{Y}$ because $\Phi$ is approximately the identity matrix, up to exponentially small diagonal corrections. To be more systematic, we can absorb these corrections inside $\delta\textbf{Y}$, and then make the remainder of the argument. By the Cauchy-Schwarz inequality, the rows of this newly defined $\delta\textbf{Y}$ also obey a bound on their norms.} which is exponentially suppressed because of \eqref{eq:suppression}. Furthermore, the variance in the distribution of the eigenvalues can also similarly be shown to be exponentially small. Thus, almost all the eigenvalues of $\textbf{Y}$ will be unaffected by the correction term $\delta\textbf{Y}$, and thus be bounded away from zero, i.e., almost all the conjugate points (which were present at small $\mu$) will get lifted. It is nevertheless true that this argument does not preclude large corrections to a small number of the eigenvalues, and thus does not completely rule out ``accidental'' conjugate points; it will be interesting to see if this can be accomplished by using more detailed properties of $\delta\textbf{Y}$. 

We emphasize that the potential remaining conjugate points discussed above are ``accidental" from the perspective of a random family of Hamiltonians in the following sense:
in quantum circuit complexity, we are concerned with families of Hamiltonians and therefore with families of conjugate points.
In the bi-invariant analysis, we found that conjugate points were very generic close to the identity, and specifically that any family of random Hamiltonians will have a family of conjugate points in the bi-invariant metric with distances from the identity set by the total spectral range.
If the entries of the Hamiltonian have mean zero and unit variance, this conjugate point family is actually moving closer to the identity as we increase $N$.
The ``accidental" conjugate points above are not generic in this way, and require some fine tuning of the matrix $\delta \textbf{Y}$.
Therefore, we do not expect them to exist in families (i.e. for arbitrary $N$), and even if we are unfortunate enough to encounter such a family, we expect that a small perturbation of the Hamiltonians will destroy them.

\end{section}

\begin{section}{Discussion}\label{sec:discussion}
In this paper, we study the quantum circuit complexity of unitary time evolution in qubit systems.   Here, complexity measures the minimum amount of ``simple'' (or $k$-local) operations needed to build the time  evolution operator $U(t) = e^{-iHt}$.  Our main tool is a geometrization of complexity in terms of geodesics on the unitary group manifold \cite{Nielsen2007}, which we study using the Euler-Arnold equation \cite{Balasubramanian2018}.  Using this approach, we directly relate complexity growth in a physical theory to its spectral properties, and thus to phenomena like chaos and integrability.   We propose the Eigenstate Complexity Hypothesis as a criterion on the energy eigenstates of the theory as a condition for linear complexity growth for exponential times, modulo global obstructions, that would be expected in chaotic dynamics.   We apply these ideas to the SYK model.  First, for $N=2$ fermions where the theory is integrable, we solve exactly and show that complexity grows linearly at initial times but then oscillates.  For large-$N$, where the SYK theory is chaotic, we show numerically that ECH is satisfied, thus giving evidence that complexity grows linearly for exponential time, as predicted by the duality of SYK theory with the physics of black holes \cite{Kitaev2015,Maldacena2016}.

Various features of the complexity plot in Fig.~\ref{fig:CExp} can be understood as arising from distinct traits of the underlying quantum system.
For example, the appearance of a plateau has nothing to do with a notion of complexity, but rather comes from competition between various geodesics on the unitary group manifold and a self-averaging effect at large-$N$, which both occur even in the bi-invariant geometry where all operators are considered simple.\footnote{Recall that a similar dip-ramp-plateau pattern in correlation functions appeared in integrable theories without disorder as studied in \cite{Balasubramanian2019}.}
On the other hand, the location in time of the start of the plateau depends strongly on what we select as ``simple'' (local) vs. ``complex'' (nonlocal) operations; for example, if all operators are considered ``simple'' (corresponding to a bi-invariant metric on the unitary group) the complexity plateau starts at a polynomial time in $N$, rather than at exponential time when only $k$-local operators are considered simple.
Similar statements apply for the complexity ramp and the length of the ramp, respectively.
Additionally, large-$N$ features like the ramp and plateau can be discovered at small $N$ (even $N=2$) by utilizing disorder-averaging which appears in, e.g., the SYK model.
However, doubly exponential features like the Poincar\'{e} recurrences of complexity will be washed out by disorder and so are only present for a single instantiation of the model at large-$N$.
The upshot of all this is that the disorder-averaging commonly employed in studies of the SYK model acts as a sort of crutch which replicates large-$N$ features. These features should properly be interpreted as the effects of self-averaging, which occur even in a single model instance as long as the Hamiltonian is chaotic. Furthermore, the qualitative features of the complexity plot are present without any notion of easy/hard (local/nonlocal) operations, but the particular time scales which appear hinge crucially on the introduction of such a notion (defined, say, through cost factors in the complexity metric on the unitary group).

\subsection{Late-time saturation}\label{sec:LTS}
From physical considerations, it is expected that for quantum systems with gravity duals, the complexity will grow linearly until some time exponential in $N$, after which it will saturate \cite{Susskind2018}.  Conceptually, in any theory with a UV and IR cutoff this saturation will occur because of the finite dimension of the group of unitary operators acting on the Hilbert space.
This saturation of complexity is expected to arise in the geometric framework when the linear geodesic on the unitary manifold from the reference operator to the time evolution operator stops being globally minimizing.   At this point other geodesics take over.  Above, we demonstrated criteria for local minimality of the linear geodesic, i.e. under what conditions we can exclude other geodesics that are deformations of the linear one.   However, since the unitary group is compact, globally there may be geodesic loops, and we have not studied these. In the simple case with a bi-invariant metric on the unitary group,  i.e. setting the cost $\mu = 0$ to take all operators to be ``simple'', we can get a glimpse of the relevant physics.\footnote{See \cite{Yang:2018nda} for a different interpretation of the physics of the bi-invariant situation.} In this case, the geodesic equation and boundary conditions are 
\begin{equation}
    \frac{d}{ds}V(s) = 0,\;\;e^{-iV(0)}= e^{-iHt}.
\end{equation}
The solutions are given by
\begin{equation}
    V_{\vec{k}}(s) = \sum_{n=1}^{2^{N/2}} (E_n t+ 2\pi k_n) \ket{n}\bra{n},
\end{equation}
where $\ket{n}$ are the energy eigenstates of the Hamiltonian, and $\vec{k} = (k_1,\cdots, k_{2^{N/2}})$ are integers which sum up to zero because of the traceless condition. Therefore, the complexity is given by
\begin{equation}\label{minComp}
\mathcal{C}(t) = \text{min}_{\vec{k}}\,\left[e^{-S}\sum_n(E_n t+ 2\pi k_n)^2\right]^{1/2},
\end{equation}
where the minimization is over all integer vectors $\vec{k}$ subject to the constraint that they sum up to zero. In Fig.~\ref{PlateauFig} we show a numerical plot\footnote{We used the function $\texttt{NMinimize}$ in \emph{Mathematica} to make this plot. 
We cannot guarantee that  numerical minimization has converged to the true global minimum, and so these  plots should be regarded as upper-bounds on the true complexity. } of $\mathcal{C}(t)$ for ten different realizations of the SYK model with $N=8,\,q= 3$ and $\mathcal{J}=1$. In all the cases,  complexity grows linearly with time for a while, but then saturates at a time of order $t\sim 2\pi/\Delta_{\text{max}}$, which recall is precisely the time when conjugate points appear in the bi-invariant case! Another important feature to note here is the saturation for each individual SYK realization, arising from different geodesics (i.e., different integer vectors $\vec{k}$) dominating the complexity at late times. In section \ref{sec:numericsN=2}, we found similar saturation behavior in the $N=2$ case after disorder averaging; at larger $N$, each individual realization seems to ``self-average'' to produce a plateau, as is evident from the sum in equation \eqref{minComp}. 

\begin{figure}[t]
\centering
\includegraphics[height=5cm]{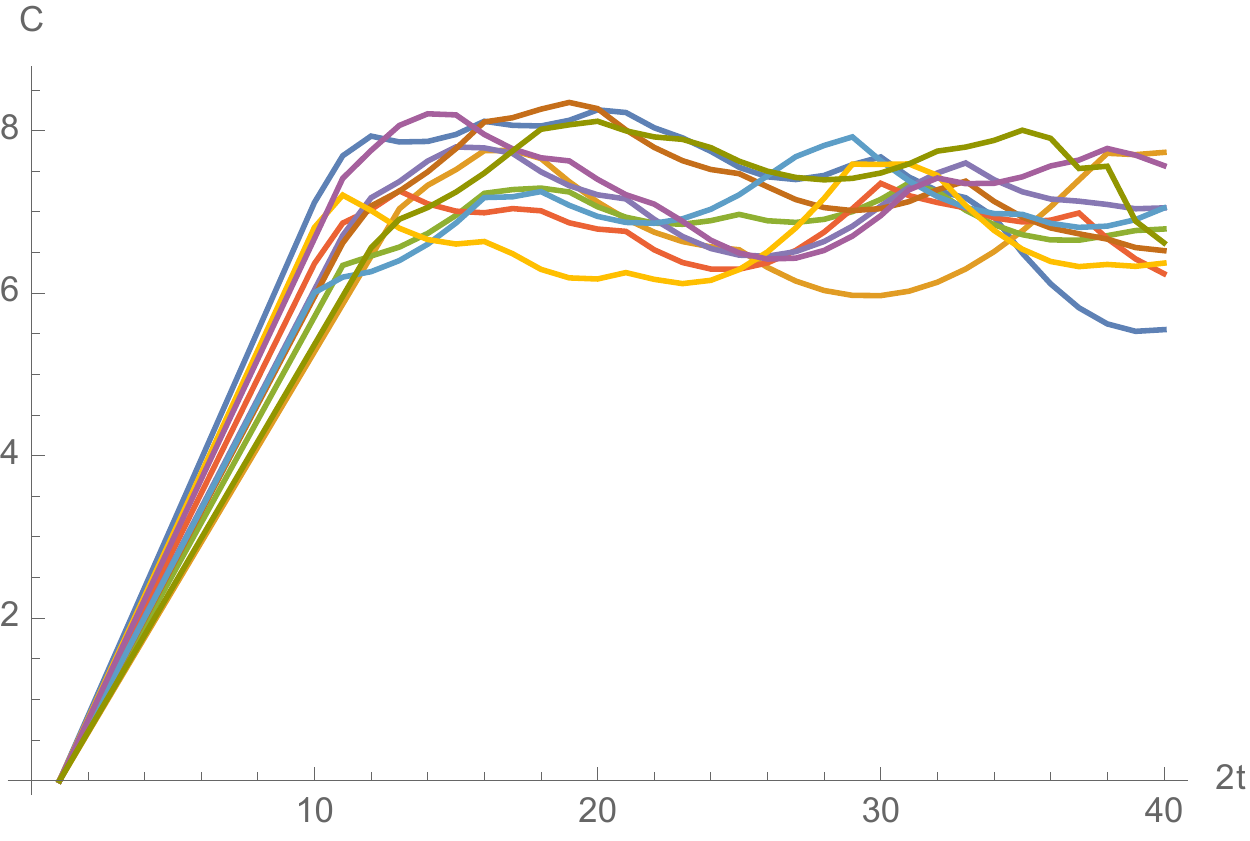}
\caption{\small{The complexity (plotted upto an overall coefficient) in the bi-invariant case ($\mu=0$) as a function of time for ten different realizations of the SYK model at $N=8,\,q=3,\,\mathcal{J}=1$.}\label{PlateauFig}}
\end{figure}

We can think of the minimization problem in equation \eqref{minComp} as being roughly equivalent to a particle moving uniformly on a $2^{N/2}$-dimensional torus $T^{2^{N/2}}$, starting from some initial point with the velocity $(E_1,\cdots, E_{2^{N/2}})$. The complexity is then simply the distance of the particle from its starting point. If the energy eigenvalues are suitably commensurate, then the distance from the starting point will grow linearly with time for some time, but then the particle will return to its origin, and this will result in an oscillating complexity. However, if the energy eigenvalues are incommensurate, then the particle will move away linearly, but will not come back to its origin in a short amount of time. Indeed, it will typically wander around the high-dimensional torus at a fixed average distance from the origin, thus leading to a saturation in the complexity. 

It would be interesting to extend this analysis to the physically interesting situation where only $k$-local operators are taken to be ``simple'' ($\mu=0$), with cost $1+\mu = 1+e^{(1-\epsilon) S}$ for all other operators.   A possible general strategy to make progress is to analyze the appearance of the complexity plateau for theories that satisfy our Eigenstate Complexity Hypothesis.

\subsection{Quantum computation}
As discussed above, to show that the linear solution is the global minimizer for an exponential time we need to globally exclude other geodesics.  In the bi-invariant geometry ($\mu = 0$, all operators regarded as ``simple''), all geodesics which reach the unitary $U = e^{-iHt}$ from the identity have initial velocity vectors equal to $\log U$.  The ambiguity in taking this logarithm gave a family of geodesics indexed by $\vec{k} \in \mathbb{Z}^{2^{N/2}}$ with $\sum_n k_n = 0$, as explained in the previous section.   Now consider some of the operators as ``non-local'' by turning on a cost factor in the metric for these directions in the unitary group. For large enough $N$ we expect all geodesics that appeared in the bi-invariant analysis other than the linear one will have nonvanishing components along the non-local directions.  Thus, when $\mu\neq 0$ these trajectories should no longer be geodesics. Perturbatively, it is obvious that their length increases with $\mu$, but a complete analysis requires a resummation of the perturbative expansion that accounts for the change in the geodesic trajectory as the metric is changed.   The goal should be to demonstrate that, at large enough $N$, all of the non-trivial geodesic loops (if they still exist when $1 + \mu \sim 2^N$) have greater length than the linear solution for any $t \sim \text{poly}(N)$, where $\text{poly}(N)$ is a polynomial of any degree.

A precise argument to this effect could be combined with our results to demonstrate a novel complexity class separation.\footnote{Hamiltonian simulation has been studied in the context of complexity classes previously in quantum computation \cite{Chuang2018}.}
This is due to a theorem of Aaronson and Susskind \cite{Aaronson2018}, who showed that the complexity class separation PSPACE $\nsubseteq$ BQP/poly is true if and only if the time evolution operator $e^{-iHt}$ in general has complexity which grows linearly with $t$ for a time greater than any polynomial in $N$.\footnote{PSPACE is the class of problems that can be solved given polynomial space and BQP is the class of problems that can be solved in polynomial time by a family of quantum circuits that is constructed by a classical algorithm in polynomial time.  Next, BQP/poly is the class of problems that can be solved in polynomial time by a family of quantum circuits, given a polynomial size string of ``advice'' for each problem size which can be used when constructing the corresponding circuit.  The advice string can be different for different problem sizes.   Here ``BQP'' stands for Bounded-error Quantum Polynomial time, where, because quantum computations are effectively probabilistic, we must require that errors occur with a probability less than some bound $\epsilon$. Finally, BQSUBEXP is the class of quantum computations that can be done in a subexponential time, $t \in O(e^{N^\alpha})$ (for all $\alpha > 0$), and BQSUBEXP/subexp is the same class but with subexponential size advice strings.} Of course, we expect such growth only for chaotic Hamiltonians, and not in integrable systems. Furthermore, following Theorem 2 in \cite{Aaronson2018}, if the geodesic loop argument outlined above can be made for exponential times (or, more precisely, for times greater than any subexponential\footnote{Note that some authors disagree on the definition of the subexponential class, effectively over whether it includes times like $2^{N^{1/3}}$ (more generally, $2^{o(n)}$) or whether it only includes times strictly less than $2^{N^\alpha}$ for all $\alpha > 0$.  Our bound on conjugate points holds for a truly exponential time $t_c \sim e^{\epsilon N}$, so the discussion of BQSUBEXP/subexp in the main text holds for whichever definition was used by \cite{Aaronson2018}.  In the previous footnote we assumed the weaker SUBEXP = $\bigcap_{\alpha > 0}$ DTIME$(2^{N^\alpha})$.}), then our our results (which show there are no conjugate points up to exponential time $t_c \sim e^{\epsilon N}$) would actually imply the even stronger statement PSPACE $\nsubseteq$ BQSUBEXP/subexp.
It would be interesting (and necessary for the aforementioned class separations to be established) to see if there is a relationship between our ECH criterion, which is central to the argument for complexity growth, and the complexity-theoretic assumption in \cite{Aaronson2018} where the time evolution step $e^{-iH}$ was taken to implement one step of a reversible computationally-universal classical cellular automaton.\footnote{It is an empirical observation in complexity theory that universality in cellular automata is not difficult to achieve; on the contrary, it is quite difficult to \textit{avoid} \cite{Neumann1966}.  We consider it very likely that a generic chaotic Hamiltonian like the SYK model implements such an automaton via its time evolution steps.}

When conjugate points exist in our analysis they can be interpreted in terms of ``fast-forwarding'' of the Hamiltonian, and of time evolution regarded as a quantum computation. 
Fast-forwarding of a Hamiltonian $H$ occurs when time evolution with respect to $H$ for a time $t$ can be simulated on a quantum computer, using a different Hamiltonian, in a time much smaller than $t$ \cite{Aharonov2017}.
General Hamiltonian simulation algorithms are well-studied in the quantum computation literature \cite{Berry2007,Childs2011,Childs2012,Berry2014,Berry2015,Aharonov2017,Aharonov2018}.
In particular \cite{Aharonov2017} shows the existence of a family of Hamiltonians (based on Shor's algorithm) where an exponential fast-forwarding does happen.  In our language, this means that there there is shorter path from the identity to the operator $e^{-iHt}$ than simply following the linear geodesic on the unitary manifold.  The existence of a conjugate point does not signal a parametrically faster algorithm, as in the definition of \cite{Aharonov2017}, but the absence of a conjugate point is certainly necessary to rule out such speedups.
Perhaps there is a connection between the existence of conjugate points (or maybe the failure of ECH) and violations of the computational time-energy uncertainty principle defined in \cite{Aharonov2017} to detect speedups.

\subsection{Quantum chaos}

We have proposed the eigenstate complexity hypothesis (ECH) as a criterion for complexity growth for exponential time, a phenomenon that we should expect in chaotic theories, but not in integrable theories. ECH states, roughly, that the off-diagonal projectors of eigenstates should have exponentially small overlap with $k$-local operators.   We have demonstrated that ECH is indeed satisfied in chaotic systems such as the SYK model. Physically, ECH is satisfied in these cases because a given projector $|m\rangle\langle n|$ has nonzero overlap with all the $e^{2S}$ operators in the $e^S$ dimensional Hilbert space, which guarantees that the overlap with a given small set of $k$-local generators must be small by unitarity (see Fig.~\ref{cDist}, appendix \ref{app:ECH}).

But what about integrable systems, such as the free Ising model
$H= -J\sum_i Z_i Z_{i+1}$?  All spin configurations are eigenstates of this model.  A spin configuration can be turned into another by the action of local raising and lowering operators at some sites and the action of any number of $Z$s.  Because of this there are $\sim e^S$ generators which will have overlaps with a given eigenstate projector with a few spin flips such as $\ket{...,1,...,0,...}\bra{...,0,...,1,...}$.  Unitarity then suggests that the overlap of this projector with any given $k$-local operator (effectively the square root of \eqref{RDef}) will be $\sim e^{-S/2}$.  This does {\it not} satisfy the ECH criterion as we stated it, but suggests there should be more refined criteria separating theories that have, e.g., O(1), O(poly) and various weaker exponential overlaps between the eigenstate projectors and the $k$-local generators. More generally, we see in Fig.~\ref{fig:p4q4} (appendix~\ref{app:ECH}) that when a system has conserved charges which act diagonally on the generators of the gate set, there are superselection sectors in the Hilbert space for the overlaps between the eigenstate projectors and the $k$-local operators.  It is plausible that this phenomenon could be shown to generally lead to ECH violation in integrable models.

All of our results were developed in the context of finite dimensional systems. These could be understood as a discretization of continuum field theories with both an IR and a UV cutoff.   It would be interesting to understand how to recover the continuum limit as the cutoffs are removed.

\subsection*{Acknowledgments}
We are grateful to Yosi Atia, Adam Bouland, Lampros Lamprou, Cl\'{e}lia de Mulatier, Daniel Ranard, G\'{a}bor S\'{a}rosi, Edward Witten, Zhenbin Yang, and Ying Zhao for useful discussions.  VB, OP, AK and MD were supported in part by the Simons Foundation through the It From Qubit Collaboration (Grant No. 38559), and also by the DOE (Contract No. FG02-05ER-41367 \& QuantISED grant DE-SC0020360). MD is supported by the National Science Foundation Graduate Research Fellowship under Grant No. DGE-1845298.
\end{section}

\begin{appendices}

\begin{section}{Majorana Fermion Basis for $\mathfrak{su}(2^{N/2})$}\label{sec:basis}
We begin with a set of Majorana fermion operators $\gamma_i$ which obey the commutation relations\footnote{Note that we are labeling gamma matrices with $i$, $j$, etc., whereas in the main text we used $a$, $b$, etc.  The reason for this is that here we reserve early alphabet letters for the binary form of base-10 integers which form an equally valid labeling of the generators that we employ in calculations.}
\begin{equation}
\{ \gamma_i , \gamma_j \} = 2 \delta_{ij} ,
\end{equation}
and also obey $\gamma_i^\dagger = \gamma_i$ (the Majorana condition).
We can interpret these objects as $2^{N/2} \times 2^{N/2}$ Hermitian matrices, and they are precisely the generalized gamma matrices of the Clifford algebra $\mathcal{C}\ell_N(\mathbb{R})$.
The basis for $\mathfrak{su}(2^{N/2})$ is constructed by taking products of $\gamma_i$ with appropriate factors of $i$ to ensure Hermiticity.
Specifically, we consider ordered products $\gamma_{i_1} \dots \gamma_{i_n}$ with $i_1 < \dots < i_n$.
To be Hermitian, such a product needs a factor of $i$ if $n(n-1)/2$ is odd.
We can now write the set of generators compactly using the set of binary strings of length $N$, $b \in \mathcal{B}_N$.
The bits of the string are $b = b_N \dots b_1$, and let $q_b$ be the number of nonzero bits in $b$.
We write
\begin{equation}
T_b = i^{{q_b}\choose{2}}\gamma_1^{b_1} \dots \gamma_N^{b_N} ,
\end{equation}
and we then have
\begin{equation}
\mathfrak{su}(2^{N/2}) = \text{span}\left( \{ T_b\ |\ b \in \mathcal{B}_N \setminus \{0\} \} \right) .
\end{equation}
We now show these generators are traceless.
By construction, the gamma matrices individually are traceless, so we have $\tr \gamma_i = 0$.
Additionally, an even number of them will be traceless since we have anticommutation and cyclicity of the trace:
\begin{equation}
\tr \gamma_{i_1} \dots \gamma_{i_{2k}} = 0.
\end{equation}
For an odd number, we use the gamma matrix construction in terms of tensor products of Pauli matrices.
That is, given a set of $N-2$ gamma matrices $\gamma_a^{(N-2)}$, we may create a set of $N$ $\gamma_i^{(N)}$ by taking
\begin{equation}
\begin{split}
\gamma_a^{(N)} & = \gamma_a^{(N-2)} \otimes \sigma_3 ,\\
\gamma_{N-1}^{(N)} & = 1 \otimes \sigma_1, \\
\gamma_N^{(N)} & = 1 \otimes \sigma_2 .
\end{split}
\end{equation}
There are four cases for $\tr \gamma_{i_1} \dots \gamma_{i_{2k+1}}$.
First, neither $\gamma_{N-1}^{(N)}$ nor $\gamma_N^{(N)}$ appear in the product.
In that case, we have $\gamma_{i_1}^{(N)} \dots \gamma_{i_{2k+1}}^{(N)} = \gamma_{i_1}^{(N-2)} \dots \gamma_{i_{2k+1}}^{(N-2)} \otimes \sigma_3^{2k+1}$, and since $\tr A \otimes B = \tr A \tr B$ and $\tr \sigma_3^{2k+1} = \tr \sigma_3 = 0$, we have the required result.
The second case is when either $\gamma_{N-1}^{(N)}$ or $\gamma_N^{(N)}$ appear in the product, but not both.
Then the final tensor factor is either $\sigma_1$ or $\sigma_2$, and similarly we have $\tr \sigma_1 = \tr \sigma_2 = 0$ so again the entire trace vanishes.
The interesting case is when we have both $\gamma_{N-1}^{(N)}$ and $\gamma_N^{(N)}$ in the product.
Then we have $\tr \gamma_{i_1}^{(N)} \dots \gamma_{i_{2k+1}}^{(N)} = - \tr \gamma_{i_1}^{(N-2)} \dots \gamma_{i_{2k-1}}^{(N-2)} \tr \sigma_1 \sigma_2 \sigma_3$.
We now repeat the argument for $\tr \gamma_{i_1}^{(N-2)} \dots \gamma_{i_{2k-1}}^{(N-2)}$, since these smaller gamma matrices have a similar tensor product structure.
Again the only interesting case is when both $\gamma_{N-3}^{(N-2)}$ and $\gamma_{N-2}^{(N-2)}$ both appear.
Following this chain, we end up with only a single gamma matrix in the first part of the tensor structure, and the trace of any single gamma always vanishes.
So we conclude
\begin{equation}
\tr \gamma_{i_1} \dots \gamma_{i_{2k+1}} = 0.
\end{equation}
Thus, all generators are traceless as desired.
\begin{equation}
\tr T_b = 0.
\end{equation}
We now turn to linear independence.
Notice first that, given generators $T_a$ and $T_b$ with $a \neq b$, we have $T_a T_b = \alpha T_c$ for some $\alpha \in \mathbb{C}$ and some $c \in \mathcal{B}_N \setminus \{0\}$.
Now assume for the sake of contradiction that we have $\sum_b \alpha_b T_b = 0$ for some constants $\alpha_b \in \mathbb{R}$.
Solve for a specific $T_a$ with nonzero coefficient and write $\alpha_a T_a =- \sum_{b \neq a} \alpha_b T_b$.
Multiply both sides by whatever multiple of $T_a$ we need to get the identity on the left hand side.
We now have $1 \propto \sum_{b \neq a} \alpha_b T_a T_b \propto \sum_{c\neq a} \beta_c T_c$.
However, if we now take the trace of both sides, the left hand side is $\tr 1 = 2^{N/2}$ but the right hand side is $\sum_c \alpha_c \tr T_c = 0$, so they cannot be proportional.
Thus, all the generators must be linearly independent.

Note that it is a basic fact of Lie algebras that the structure constants ${f_{ab}}^c$ are fully antisymmetric since we have chosen a basis in which $\tr T_a T_b \propto \delta_{ab}$.
This can be seen by noticing $\tr T_a [T_b,T_c] = \tr T_a {f_{bc}}^d T_d \propto {f_{bc}}^d \delta_{ad} = {f_{bc}}^a$, and also by cyclicity of the trace we have
\begin{align*}
\tr T_a [T_b,T_c] & = \tr T_a T_b T_c - \tr T_a T_c T_b \\
& = \tr \tr T_a T_b T_c - \tr T_b T_a T_c \\
& = \tr [T_a,T_b] T_c \\
& = {f_{ab}}^d \tr T_d T_c \\
& \propto {f_{ab}}^d \delta_{dc} \\
& = {f_{ab}}^c ,
\end{align*}
therefore we have
\begin{equation}
{f_{bc}}^a = {f_{ab}}^c ,
\end{equation}
and this combined with the fact that ${f_{ab}}^c = -{f_{ba}}^c$ implies full antisymmetry.
Now recall $q_i$ is the number of nonzero bits in the binary expression of index $a$, i.e. $q_a$ is the number of fermions appearing in generator $a$.
Let $a \oplus b$ be the bitwise ``exclusive-or" of $a$ and $b$.
Let $a \land b$ be the bitwise ``and" of $a$ and $b$.
Then a lengthy calculation shows 
\begin{equation}
{f_{ab}}^c \neq 0 \Leftrightarrow a \oplus b = c,\ q_a q_b + q_{a\land b} \equiv 1 \mod 2 .
\label{eq:structurecondition}
\end{equation}
Furthermore, the magnitude of any nonzero structure constant is precisely $|{f_{ab}}^c| = 2$.
The exact sign is more difficult to determine with simple calculations, but it will not be so important for our analysis.
One may wonder how this description of the structure constants is consistent with the properties we claimed before.
For instance, if $a \oplus b = c$ and $q_a q_b + q_{a\land b}$ is odd, do we have (by total antisymmetry) $b \oplus c = a$ and $q_b q_c + q_{b \land c}$ odd?
Indeed we do, by properties of $\oplus$ and $\land$:
\begin{align*}
a \oplus b & = c \\
a \oplus (a\oplus b) \oplus c & = a \oplus c \oplus c \\
b \oplus c & = a .
\end{align*}
Another computation shows the parity of $q_a q_b + q_{a\land b}$ matches that of $q_b q_c + q_{b \land c}$.
\begin{align*}
q_a q_b + q_{a\land b} & \equiv 1 \mod 2 \\
q_a q_b + q_b q_c + q_{a\land b} + q_{b\land c} & \equiv 1 + q_b q_c + q_{b\land c} \mod 2 \\
(q_a + q_c) q_b + q_{a\land b} + q_{b\land c} + 1 & \equiv q_b q_c + q_{b\land c} \mod 2 \\
(q_a + q_{a \oplus b}) q_b + q_{a\land b} + q_{b \land (i\oplus b)} + 1 & \equiv q_b q_c + q_{b\land c} \mod 2 \\
(q_a + q_a + q_b - 2q_{a \land b}) q_b + q_{a\land b} + q_b - q_{a\land b} + 1 \equiv q_b q_c + q_{b \land c} \mod 2 \\
q_b^2 + q_b + 1 \equiv q_b q_c + q_{b\land c} \mod 2 \\
q_b (q_b + 1) + 1 \equiv q_b q_c + q_{b\land c} \mod 2 \\
1 \equiv q_b q_c + q_{b\land c} \mod 2 .
\end{align*}
This serves as a consistency check on our condition \eqref{eq:structurecondition}.
Notice that the condition \eqref{eq:structurecondition} acts as a nontrivial selection rule for which velocities can appear in the Euler-Arnold equation \eqref{eq:eulerarnold}.
Let us see how this works.
For a local direction $a$, in order for $V^b V^c$ to appear on the right hand side of \eqref{eq:eulerarnold}, we must have (without loss of generality) $q_b \leq k$ and $q_c > k$ for some choice of $k$-locality of the gate set.
Then, since $q_c = q_a + q_b - 2q_{a\land b}$, we must have $k > 2q_{a\land b}$.
For $k=2$, this implies $q_{a\land b} = 0$.
Combining with $q_a q_b + q_{a\land b} \equiv 1 \mod 2$, we see $q_a q_b \equiv 1 \mod 2$.
This is quite a nontrivial condition; na\"{i}vely we may have imagined in a $k=2$ model that there exist nonzero commutators between generators $a$ and $b$ that had $q_a = q_b = 2$ and $q_{a\land b}=0$, but this cannot be the case because then $q_a q_b \equiv 0 \mod 2$.
Similarly, we might have expected nonzero commutators for $q_a = 1$ and $q_b = 2$ yielding $q_c = 3$ with $q_{a\land b} = 0$, but this also does not occur by the selection rules \eqref{eq:structurecondition}.
The conclusion of this analysis is that, in a $k=2$ model, the local velocities evolve via $\frac{dV^\alpha}{ds} = 0$.

\end{section}

\section{Conjugate points in perturbation theory}\label{app:ConjugatePoints}
In the main text it was argued that the bi-invariant metric (with $c=\bar{c}=1$) has conjugate points at $t_{mn}^\star  = \frac{2\pi}{\Delta_{mn}}\mathbb{Z}$, where $\Delta_{mn} = (E_m-E_n)$ are differences between energy eigenvalues of the Hamiltonian. In this appendix, we will perturbatively track the behavior of these conjugate points when we turn on an infinitesimal cost factor $\bar{c} =1+ \epsilon$ along the heavy directions. Recall that the equation for the Jacobi field (i.e., the Euler-Arnold equation linearized around the linear geodesic)  takes the form
\begin{equation}\label{J01}
c\frac{d}{ds}\delta V_{L}(s)= -it(\bar{c} -c) \left[H, \delta V_{NL}(s)\right]_L,
\end{equation}
\begin{equation}\label{J02}
\bar{c}\frac{d}{ds}\delta V_{NL}(s)= -it(\bar{c}-c) \left[H, \delta V_{NL}(s)\right]_{NL}.
\end{equation}
Here the subscripts $L$ and $NL$ stand for projections of the corresponding operators along local and non-local directions respectively. We wish to check whether there exists some initial boundary condition $\delta V(0)$ and some value of $t$, such that
\begin{equation}\label{disp}
    U^{-1}\delta U(1) = \int_0^1 ds e^{itsH}\delta V(s) e^{-istH} = 0.
\end{equation}
If so, then the corresponding value of $t$ constitutes a conjugate point along the original linear geodesic $U_{\text{linear}}(s) = e^{-ist H}$, at which point the linear geodesic stops being globally minimizing. 

We take $c=1$ and $\bar{c} = 1 + \epsilon$, and expand
\begin{equation}
    \delta V(s) = \delta V^{(0)}(s) + \epsilon \delta V^{(1)}(s) +\epsilon^2 \delta V^{(2)}(s) \cdots.
\end{equation}
Expanding equations \eqref{J01} and \eqref{J02} at zeroth order, we find that $\delta V^{(0)}(s)=\delta V^{(0)}(0)$, i.e., it is $s$-independent. Indeed, this is the bi-invariant Jacobi field, which gives the family of conjugate points at 
\begin{equation}
    t_{(m,n)}^\star  = \frac{2\pi}{\Delta_{mn}}\mathbb{Z}, \;\;\;\delta V^{(0)}(0) = z |m\rangle \langle n| +\bar{z}|n\rangle \langle m|, 
\end{equation}
for any pair of distinct eigenvalues $m$ and $n$ of the Hamiltonian with $\Delta_{mn} = (E_m-E_n)$, and any non-zero complex number $z$. We can set the absolute value of $z$ to one by choice of normalization, but its phase is not determined at this order. This implies a two-fold ``degeneracy'' in all the bi-invariant conjugate points. 

At first order in $\epsilon$, we find the equations
\begin{equation}\label{J1}
\frac{d}{ds}\delta V^{(1)}_{L}(s)= -it \left[H, \delta V^{(0)}_{NL}(0)\right]_L,
\end{equation}
\begin{equation}\label{J2}
\frac{d}{ds}\delta V^{(1)}_{NL}(s)= -it \left[H, \delta V^{(0)}_{NL}(0)\right]_{NL},
\end{equation}
which can be solved to obtain
\begin{equation}
    \delta V^{(1)}(s) =\delta V^{(1)}(0) -ist \left[H, \delta V^{(0)}_{NL}(0)\right].
\end{equation}
From equation \eqref{disp}, we then find that the final displacement at the present order is given by
\begin{eqnarray}\label{disp1}
    \langle m'| U^{-1}\delta U(1)|n'\rangle  &=& \int_0^1 ds e^{its\Delta_{m'n'}}\left[\langle m'| (\delta V^{(0)}(0) + \epsilon\delta V^{(1)}(0))|n'\rangle \right.\nonumber\\
    &-&\left. i\epsilon st\Delta_{m'n'} \langle m'| \delta V^{(0)}_{NL}(0)|n'\rangle \right]+O(\epsilon^2)\\
    &=&\phi_{m'n'}(t)\langle m'|\delta V^{(0)}(0)+\epsilon \delta V^{(1)}(0)|n'\rangle -\epsilon t\partial_t\phi_{m'n'}(t) \langle m'| \delta V^{(0)}_{NL}(0)|n'\rangle+O(\epsilon^2)\nonumber, 
\end{eqnarray}
where found it more convenient to write the matrix elements of $U^{-1}\delta U(1)$ in the energy eigenstates $|m'\rangle,\,|n'\rangle$, and we have defined the function
$$\phi_{m'n'}(t)= \frac{e^{it\Delta_{m'n'}}-1}{it\Delta_{m'n'}}.$$
Let us now return to the conjugate points $t_{mn}^\star  = \frac{2\pi}{\Delta_{mn}}\mathbb{Z}$ we had obtained at zeroth order. From equation \eqref{disp1}, it is clear that their locations have now moved at linear order in $\epsilon$, which we can keep track of systematically in perturbation theory:
\begin{equation}
  t^{\star}_{mn} = t^{\star,(0)}_{mn}+ \epsilon \delta^{(1)} t^\star_{mn} + \epsilon^2 \delta^{(2)} t^{\star}_{mn}+\cdots,
\end{equation}
where $t^{\star,(0)}_{mn}=\frac{2\pi}{\Delta_{mn}}\mathbb{Z}$ denotes the bi-invariant conjugate points. Substituting this expansion in equation \eqref{disp1} and demanding that the result vanish for $m'=m$ and $n'=n$,\footnote{Of course, we should also demand that the matrix elements of $U^{-1}\delta U(1)$ vanish for $m\neq m'$ and $n\neq n'$; these constraints partially determine $\delta V^{(1)}(0)$.} we deduce the shifts in the conjugate points:
\begin{equation}\label{CP1}
    \delta^{(1)} t^{\star}_{mn} = t^{\star,(0)}_{mn}\frac{\langle m | \delta V^{(0)}_{NL}(0)| n \rangle}{\langle m | \delta V^{(0)}(0)| n \rangle} = t^{\star,(0)}_{mn}\left(1-\frac{\langle m | \delta V^{(0)}_{L}(0)| n \rangle}{\langle m | \delta V^{(0)}(0)| n \rangle}\right) .
\end{equation}
Note that the numerator on the right hand side involves the projection of $\delta V^{(0)}(0)$ to the non-local subspace, which makes the above expression somewhat non-trivial to evaluate. Importantly, however, $\delta V^{(0)}(0)$ at the zeroth order was only determined up to an arbitrary complex number $z$:
\begin{equation}
\delta V^{(0)}(0) = z |m\rangle \langle n| +\bar{z}|n\rangle \langle m|.
\end{equation}
Requiring that our expression for $\delta^{(1)} t_{mn}^{\star}$ is real determines precisely two possible choices of $z$, which we may call $z_{\pm}$. (Actually, only the phase of $z$ is determined. The absolute value can be set to one by choice of normalization.) Corresponding to these two fixed numbers $z_{\pm}$, we then have the two conjugate points at their respective locations $t^{\star,(0)}_{mn}+ \epsilon \delta^{(1)}t^{\star}_{mn}(z_{\pm})$, given by \eqref{CP1}. Therefore, we find that at linear order in $\epsilon$, the two-fold ``degeneracy'' in conjugate points splits. Nevertheless, they continue to exist and we have tracked their locations at $O(\epsilon)$ above. 

This analysis can be repeated order by order in perturbation theory to determine the location of conjugate points. In the main text, it was shown that for the cost factor $\bar{c}=1+\mu$, the conjugate points move to $t\sim t^{(0)}(1+\mu)$, assuming ECH. We see that the perturbative results derived here are consistent with this. In particular, we reproduce the formula in the main text if we drop the $\delta V^{(0)}_L$ term in equation \eqref{CP1}. 

\section{Some more details on ECH}
\label{app:ECH}
Here we provide some more numerical evidence for ECH in the SYK model. First, let us consider writing a generic off-diagonal eigenstate projector $\ket{m}\bra{n}$ in the SYK model in terms of the generators $T_a=(T_{\alpha}, T_{\dot\alpha})$ consisting of products of fermions:
\begin{equation}
   \ket{m}\bra{n}= \frac{1}{2^{N/2}}\sum_a c_aT_a,\;\; c_a = \langle n|T_a|m\rangle, 
\end{equation}
where $a$ runs over all the directions, easy and hard. 
\begin{figure}[t]
    \centering
    \begin{tabular}{c c}
      \includegraphics[height=4.5cm]{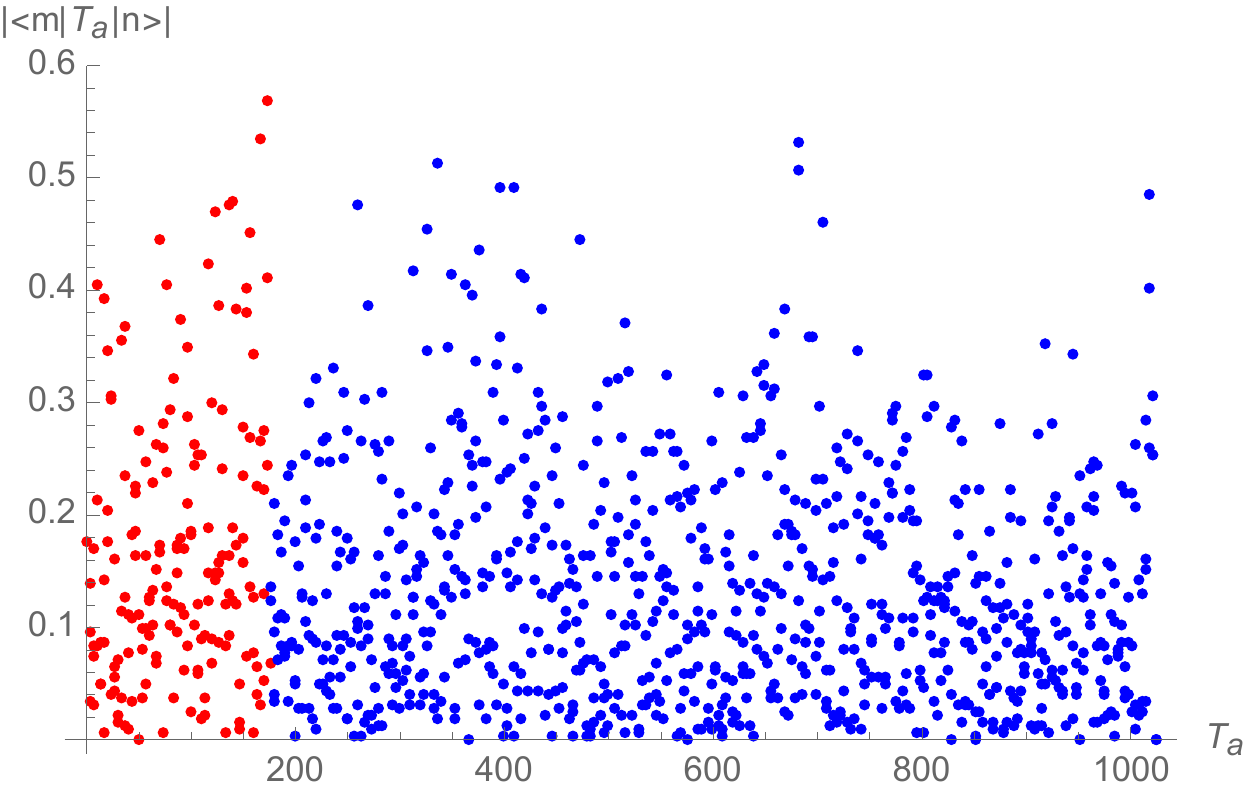}    &  \includegraphics[height=4.5cm]{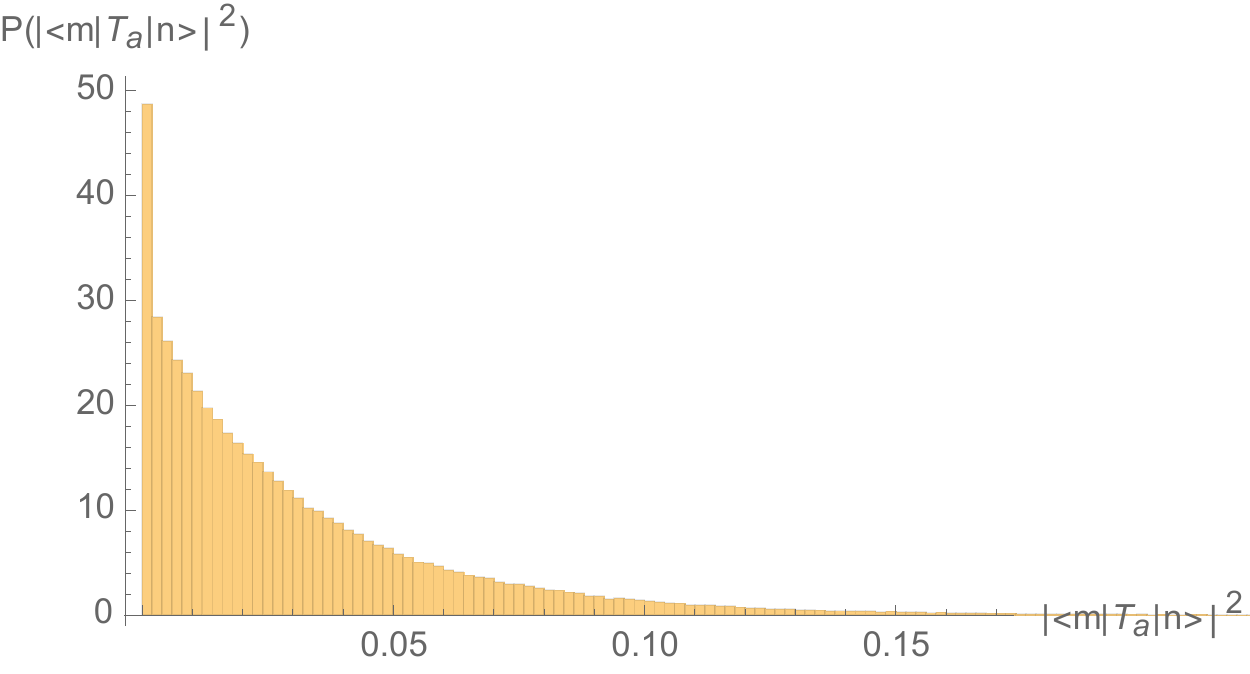}
    \end{tabular}
    \caption{\small{(Left) The absolute values of the coefficients $c_a$ for all the generators $T_a$, for a typical projector $\ket{m}\bra{n}$ in the SYK model. Red dots are the easy generators while blue dots are the hard ones. Here $N=10, k=3,q=3$. (Right) The probability distribution of $|c_a|^2$ for all $a,m,n$. }\label{cDist}}
\end{figure}
\begin{figure}[t]
    \centering
    \includegraphics[height=4.5cm]{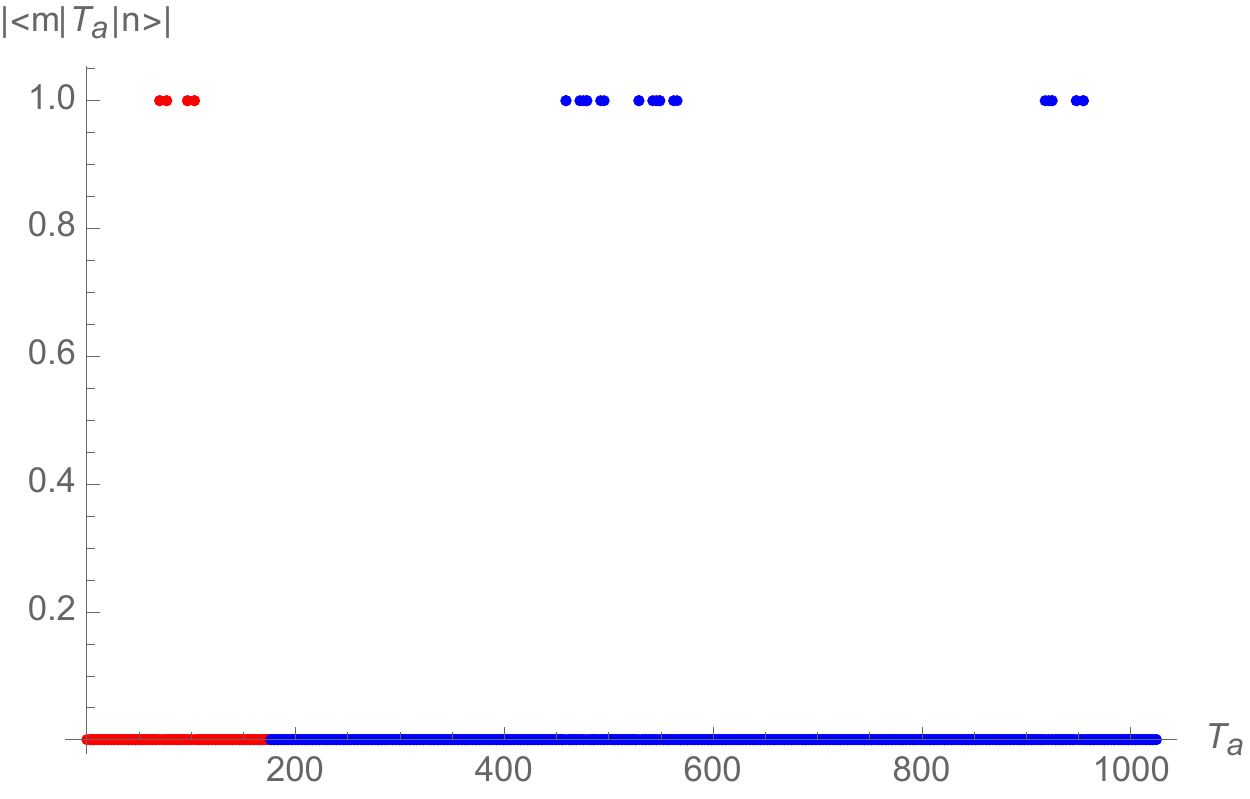}    
    \caption{\small{The absolute values of the coefficients $c_a$ for all the generators $T_a$ for a typical projector $\ket{m}\bra{n}$ of the Hamiltonian $H=i\psi_1\psi_2$. Red dots are the easy generators while blue dots are the hard ones. Here $N=10, k=3$. }\label{fig:cDistInt}}
\end{figure}
We can get some heuristic understanding of why ECH is true in the SYK model by looking at the distribution of the $c_a$. We see from the left panel of Fig.~\ref{cDist} that the $c_a$ are more or less uniformly distributed over all the $e^{2S}$ generators. Since $R_{mn}$ is the weight in the easy directions, the uniformity in the distribution of $c_a$ implies that $R_{mn}$ will be proportional to the number of easy directions divided by the total number of directions, which is precisely what ECH requires. A related comment is that if we build the distribution of the $c_a$s by pooling together these coefficients for all choices of $m$ and $n$, then we find a distribution with an exponential tail (see the right panel of Fig.~\ref{cDist}). Since the tail is exponential, and the $R_{mn}$s correspond to normalized sums over the easy coefficients, we expect the distribution over $R_{mn}$s to be Gaussian in the large-$N$ limit, by the central limit theorem. This is consistent with the distribution in Fig.~\ref{ECH}. In Fig.~\ref{fig:cDistInt} we have shown the distribution of the $c_a$ for a typical projector of an integrable Hamiltonian $H=\psi_1\psi_2$. Note that in this case, the overlaps are distributed in a much smaller subset of the generators. Nevertheless, there seem to be overlaps with about $e^S$ generators (as opposed to $e^{2S}$ in the SYK model), suggesting a milder suppression of $R_{mn}$. 

\begin{figure}
    \centering
    \includegraphics[height=4.5cm]{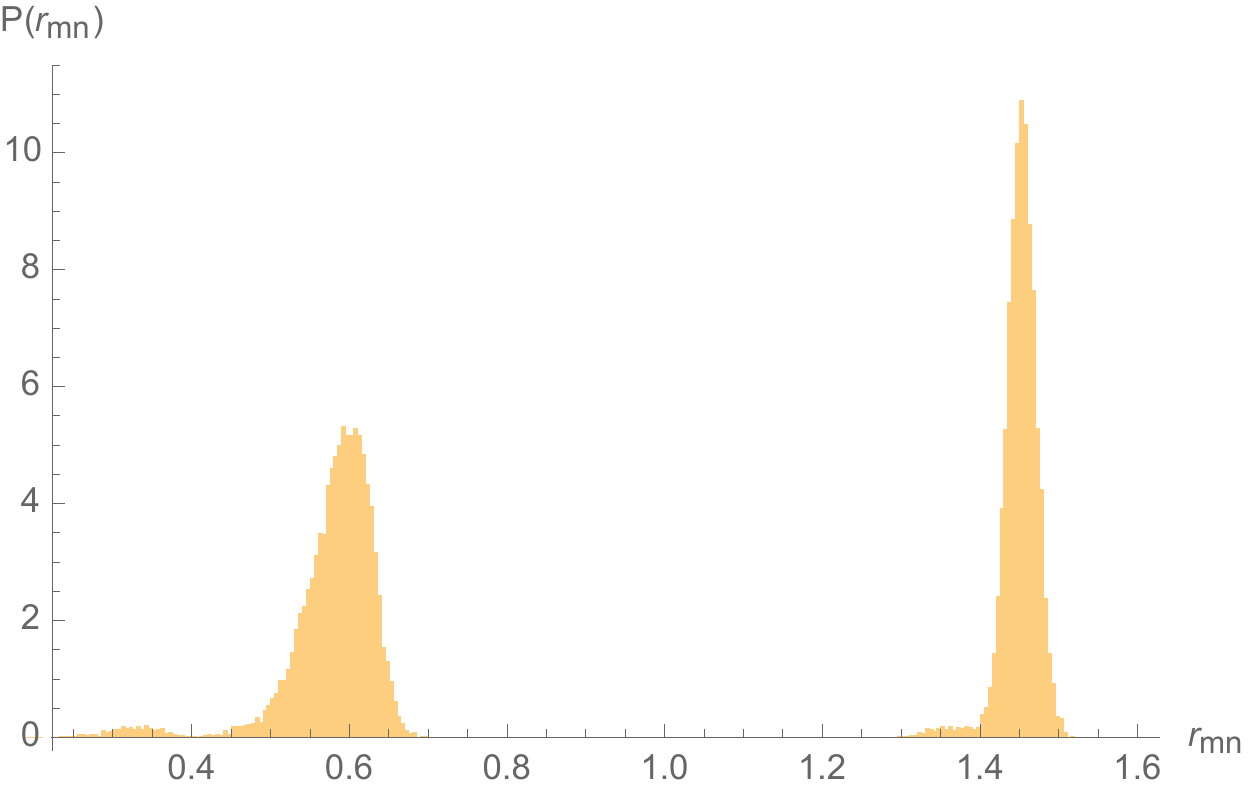}
    \caption{\small{For $k=4,q=4$ and $N=12$, the distribution of $r_{mn}$s splits into two distributions corresponding to the bosonic (right) and fermionic (left) energy eigenstate projectors.}\label{fig:p4q4}}
    \label{fig:my_label}
\end{figure}

We have mainly focused on $k=3$, $q=3$ in our presentation. But similar results also apply to $k=4$, $q=4$, with slight modifications. The main novelty is that for $q=4$, the Hamiltonian has a fermion-number symmetry. As a consequence, the eigenstates of the Hamiltonian carry an extra quantum number, namely the fermion number which acts diagonally on the generators involving products of fermions. This means that the off-diagonal projectors $\ket{m}\bra{n}$ are of two types: 1. ``Bosonic'' or fermion number preserving, 2. ``Fermionic'' or fermion number reversing. As a consequence of this, the distribution of $r_{mn}$s in this case splits into two well-localized distributions, see Fig.~\ref{fig:p4q4}. Since the fermionic projectors cannot have any overlap with the four-fermion operators in the easy part of the Lie algebra, the corresponding $r_{mn}$s are slightly suppressed (their distribution has moved to the left). On the other hand, since the average is constrained to one, this forces the bosonic $r_{mn}$s to be slightly enhanced (their distribution has moved to the right). However, these effects are polynomial in $N$, and do not affect the overall exponential suppression of all the $R_{mn}$s. 

\end{appendices}

\bibliographystyle{JHEP}
\bibliography{SYKbib}

\end{document}